\documentstyle[a4,11pt]{article}

\newcommand{\sect}[1]{\setcounter{equation}{0}\section{#1}}
\newcommand{\subsect}[1]{\subsection{#1}}

\newcommand{\vs}[1]{\rule[- #1 mm]{0mm}{#1 mm}}

\newcommand{\lbl}[1]{\label{eq:#1}}
\newcommand{\rf}[1]{(\ref{eq:#1})}
\newcommand{\nn}{\nonumber}

\newcommand{\be}{\vs{2}\large\begin{equation}}
\newcommand{\ee}{\\[2mm]\end{equation}\normalsize}
\newcommand{\cee}{\\[2mm]\]\normalsize}
\newcommand{\cbe}{\vs{2}\large\[}
\newcommand{\bea}{\large\begin{eqnarray}}
\newcommand{\ena}{\end{eqnarray}\normalsize}
\newcommand{\nnbea}{\large\begin{eqnarray*}}
\newcommand{\nnena}{\end{eqnarray*}\normalsize}

\newcommand{\lp}{\left(}
\newcommand{\rp}{\right)}
\newcommand{\tld}{\widetilde}

\newcommand{\sm}[2]{\textstyle{\frac{#1}{#2}}\displaystyle}
\newcommand{\shalf}{\textstyle{\frac{1}{2}}\displaystyle}

\newcommand{\bz}{{\bar{z}}}
\newcommand{\bw}{{\bar{w}}}
\newcommand{\zt}{{\zeta}}
\newcommand{\om}{{\omega}}
\newcommand{\si}{{\sigma}}

\newcommand{\vz}{{v_\bz}^z}
\newcommand{\vzz}{{v_\bz}^{zz}}

\newcommand{\vb}{v_\bz}

\newcommand{\vro}{{v^\rho}_\bz^{\ \zt}}
\newcommand{\vmi}{{v^{\scriptscriptstyle \ominus}}_\bz^{\ \zt}}
\newcommand{\vpl}{{v^{\scriptscriptstyle \oplus}}_\bz^{\ \zt}}
\newcommand{\wro}{W^\rho_{3/2}}
\newcommand{\wmi}{W^{\scriptscriptstyle \ominus}_{3/2}}
\newcommand{\wpl}{W^{\scriptscriptstyle \oplus}_{3/2}}
\newcommand{\cro}{c^{\rho \, \zt}}
\newcommand{\cmi}{c^{{\scriptscriptstyle \ominus} \zt}}
\newcommand{\cpl}{c^{{\scriptscriptstyle \oplus} \zt}}

\newcommand{\cz}{{{\chi_{}}_{}}_z}
\newcommand{\cbz}{{{\chi_{}}_{}}_\bz}

\newcommand{\lz}{\lambda_z}
\newcommand{\lzz}{\lambda_{zz}}
\newcommand{\lbzz}{\lambda_{\bz z}}

\newcommand{\Lzz}{\Lambda_{zz}}

\newcommand{\prt}{\partial}
\newcommand{\bp}{\bar{\partial}}
\newcommand{\dz}{\partial_z}
\newcommand{\dbz}{\partial_\bz}

\newcommand{\al}{\alpha}
\newcommand{\Ga}{\Gamma}

\newcommand{\anom}{\mbox{\huge $a$}}

\newcommand{\slt}{$SL(2)$ }
\newcommand{\sltr}{$SL(3)$ }

\newcommand{\cem}{\hspace{1cm}}

\newcommand{\ca}{{\mbox{\boldmath $\cal A$}}}
\newcommand{\cqb}{{\mbox{\boldmath $\cal Q$}}}
\newcommand{\cd}{{\cal D}}
\newcommand{\cf}{{\cal F}}
\newcommand{\cg}{{\cal G}}

\newcommand{\cv}{{\cal V}}
\newcommand{\cw}{{\cal W}}
\newcommand{\2}{\Delta^{(2)}}
\newcommand{\3}{\Delta^{(3)}}
\newcommand{\5}{\Delta^{(5)}}

\def\ie{{\it i.e.\ }}

\font\twelve=cmbx10 at 12pt
\font\ten=cmbx10 at 12pt

\renewcommand{\thefootnote}{\fnsymbol{footnote}}
\begin{document}

\begin{titlepage}

\begin{center}

{\ten Centre de Physique Th\'eorique}\footnote{Unit\'e Propre de
Recherche 7061} {\ten - CNRS - Luminy, Case 907}

{\ten F-13288 Marseille Cedex 9 - France }

\vspace{2 cm}

{\twelve {\large {$\cal W$}}-GAUGE STRUCTURES AND THEIR ANOMALIES:
        \vspace{3 mm} \\ AN ALGEBRAIC APPROACH}

\vspace{1 cm}
\setcounter{footnote}{0}
\renewcommand{\thefootnote}{\arabic{footnote}}

{\bf Daniela GARAJEU, Richard GRIMM}
\footnote{and University of Aix-Marseille II}{\bf, Serge LAZZARINI}$^{\ 1,}$
\footnote{Present address: University of Amsterdam, Institute for
Theoretical Physics, Valckenierstraat 65, NL-1018 XE Amsterdam,
The Netherlands.}

\vspace{7mm}

{\bf Abstract}

\end{center}

Starting from flat two-dimensional gauge potentials we
propose the notion of $\cw$-gauge structure in terms of a
nilpotent BRS differential algebra. The decomposition of the
underlying Lie algebra with respect to an \slt subalgebra is
crucial for the discussion conformal covariance, in
particular the appearance of a projective connection.
Different \slt embeddings lead to various $\cw$-gauge structures.

We present a general soldering procedure which allows to express
zero curvature conditions for the $\cw$-currents
in terms of conformally covariant differential operators acting on the
$\cw$ gauge fields and to obtain, at the same time,
the complete nilpotent BRS differential
algebra generated by $\cw$-currents, gauge fields and the
ghost fields corresponding to $\cw$-diffeomorphisms.
As illustrations we treat the cases of \slt itself
and to the two different \slt embeddings in \sltr, {\em viz.} the
$\cw_3^{(1)}$- and $\cw_3^{(2)}$-gauge structures, in some detail.
In these cases we determine algebraically $\cw$-anomalies as solutions
of the consistency conditions and discuss their Chern-Simons origin.

\vspace{1 cm}

\noindent November 1994

\noindent CPT-94/P.3078

\medskip

\noindent anonymous ftp or gopher: cpt.univ-mrs.fr

\end{titlepage}


\tableofcontents

\sect{INTRODUCTION}

\indent

\indent

$\cw$-symmetry \cite{BS93} intertwines internal symmetries with
space-time symmetries in two dimensions. A dynamical realization of
$\cw$-symmetry arises in the reduction of WZNW theories to Toda
field theory \cite{BFO*90},\cite{FOR*92a}. In this approach, the
original set of Lie algebra valued Kac-Moody currents of the WZNW
theory is reduced to a set of $\cw$-currents which are primary conformal
fields of well-defined (integral or half-integral) conformal
weights. An important ingredient in this construction is the
identification of a $SL(2)$ subalgebra of the Lie algebra
which underlies the WZNW theory: the remaining generators
are then arranged in irreducible representations
with respect to the $SL(2)$ subalgebra.

In general \cite{Dyn57}, \cite{Kos59},
for a given Lie algebra ${\cal G}$ there are several
possibilities to identify such a $SL(2)$ subalgebra
\cite{BTvD91},\cite{FOR*92},\cite{FRS93},\cite{FPGMG94}.
Different embeddings lead to different $\cw$-structures.
The first examples were found for the case of \sltr by
Zamolodchikov \cite{Zam85} and by Polyakov \cite{Pol90} and
Bershadsky \cite{Ber91}, the $\cw_3^{(1)}$- and $\cw_3^{(2)}$-algebras.
In any case, the $\cw$-currents correspond to the highest weight
generators in the $SL(2)$ decomposition. They are all conformally
covariant tensors except for the one in the $SL(2)$ subalgebra
itself, which behaves as a projective connection
(a property shared by the energy-momentum tensor in
two-dimensional conformal field theory \cite{Laz90}).
All the currents in these reduced WZNW theories
are holomorphic quantities, the holomorphicity conditions can be
understood in terms of zero-curvature conditions,
reflecting the integrability properties of the Toda
theories \cite{LS83} in terms of a Lax-pair formulation.
The theory which results from the reduction
procedure exhibits $\cw$-symmetry, the $\cw$-transformations are
identified as the residual gauge transformation which survive
the reduction of the original theory.

Another dynamical context in which $\cw$-symmetry arises is
$\cw$-gravity (reviews and references may
be found in \cite{Hul93} or \cite{dB93}).
These theories are conceived as generalizations of usual induced
gravity where the energy-momentum tensor couples to the metric and its
conservation is spoiled by the conformal anomaly. The most convenient
parametrization for this system is that where the metric (resp.
moving frame) is described by the Beltrami
differentials and the conformal factor (resp. an
additional chiral Lorentz factor for the frame).
In this factorized formulation
the conformal anomaly is given in terms of a
covariantly chiral third order differential operators acting on
the Beltrami differential (see \cite{Laz90} for a review and references).
The same differential operator arises in the second hamiltonian structure
of the KdV hierarchy. Integration of the conformal anomaly gives rise to
induced gravity.

In the case of $\cw$-gravity one imagines that the $\cw$-currents,
which are considered as covariant higher conformal spin
generalizations of the energy-momentum tensor, couple to certain
$\cw$-gauge fields which in turn are considered to be
generalizations of the Beltrami differentials (hence the notions
$\cw$-frame and $\cw$-diffeomorphisms).
In a conformal quantum field theory which realizes $\cw$-symmetry
one expects then that the $\cw$-currents are no longer holomorphic
quantities, the Ward identities arising from $\cw$-transformations
are anomalous. These questions have been addressed in
\cite{BFK91},\cite{DHR90},\cite{GLM91},\cite{OSSvN92}),
in particular detail for the
case of the $\cw$-structures pertaining to $SL(3)$. In these papers
special emphasis was on possible interpretations of $\cw$ Ward identities
in terms of zero curvature conditions and in relation with
integrable hierarchies and their Lax pair formulation.

On the other hand, as it is well known, anomalies must satisfy the
Wess-Zumino consistency conditions, and one can define anomalies
algebraically as nontrivial solutions to these consistency
conditions. The most powerful and elegant formulation of this
approach is in terms of the BRS differential algebra of gauge and
ghost fields (and matter fields as well).

It is this purely algebraic attitude that will be pursued in this
paper: we propose the notion of $\cw$-gauge structure in terms of
differential BRS algebra generated by $\cw$-currents, $\cw$-gauge
fields and $\cw$-ghost fields, with nilpotent operation of the exterior
space-time derivative and the BRS operator. This $\cw$-gauge structure
is obtained from a generalized zero curvature formulation. In
turn, the zero curvature condition can be viewed as a
compatibility relation for covariantly constant fields, which
will be included in our BRS analysis as well. With this algebraic
formalism at hand, in particular the nilpotent BRS differential
algebra, one can search for nontrivial solutions
of the  consistency equations.

The paper is organized as follows:
in {\it chapter 2}, based on \cite{AG92},
we present the general framework and define our
notations, in {\it chapter 3} we treat in detail the case of $SL(2)$
itself, and the Chern-Simons origin of the anomaly. {\it Chapter 4}
is devoted to the presentation of the
two different $\cw$-gauge structures deriving from
$SL(3)$, including, in each case, the structure of matter fields
and the algebraic construction of the anomalies as solutions
of consistency conditions and {\it Chapter 5} contains some concluding
remarks of more conceptual nature concerning the issue of
$\cw$-geometry.

\indent

\sect{GENERAL STRUCTURE}

\indent

\subsect{Gauge potentials, ghosts and their BRS structure}

\indent

For a given simple Lie algebra $\cg$ we consider a decomposition of the
set of generators with respect to some \slt subalgebra. In general
there are several posibilities to identify such a subalgebra and
therefore different decompositions, as for instance for $SL(3)$ where
the two different decompositions correspond to the two different
$\cw_3$-algebras of Zamolodchikov and of Polyakov and Bershadsky.
For a given decomposition of the Lie algebra $\cg$ we denote the
generators of the \slt subalgebra by $L_k$, with $k=-1,0,+1$
and commutators
\be [L_k,L_l] \ = \ (k-l) L_{k+l}. \ee
The remaining generators are arranged in irreducible representations
with respect to this subalgebra, they will be denoted
$T_{\rho \ k}^{\ \ a}$. The index $a \geq -1$, integer or half-integer,
characterizes the representation (spin $a+1$) and $k$ runs from
$-a-1$ to $a+1$ in integer steps. Finally the index $\rho$ serves
to distinguish different copies of the same spin which may occur
in the decomposition (for instance, in the second $SL(3)$ decomposition
two spin $1/2$ occur, carrying different hypercharge).

For the commutators of the \slt generators with the remaining ones
$L_0$ measures as usual the third component of the spin,
\be \left[L_0,T_{\rho \ k}^{\ \ a}\right] \ = \ -k \, T_{\rho \ k}^{ \ \
a}, \ee while $L_-$ and $L_+$ act as step operators, we define
\bea
\left[L_-,T_{\rho \ k}^{\ \ a}\right]
&=& \si_{\rho \ k}^{- \ a} \, T_{\rho \ k-1}^{\ \ a}, \\
\left[L_+,T_{\rho \ k}^{\ \ a}\right]
&=& \si_{\rho \ k}^{+ \ a} \, T_{\rho \ k+1}^{\ \ a},
\ena
with structure constants $\si_{\rho \ k}^{\pm \ a}$ in some
convenient normalization.
The remaining commutation relations are parametrized as
\footnote{in the general case, the \slt decomposition makes it necessary
to use this triple index notation, which we hope not to be too
confusing, in particular for the structure constants
$\si_{\rho \ k}^{- \ a} \, T_{\rho \ k-1}^{\ \ a}$} \be
\left[T_{\rho \ k}^{\ \ a},T_{\sigma \ l}^{\ \ b}\right] \ = \
\si_{\rho k \ \sigma l \ \ \ c}^{\ \, a \ \ \, b \ \tau m} \, T_{\tau  \
m}^{\ \ c} + \si_{\rho k \ \sigma l}^{\ \, a \ \ \, b \ m} \, L_m. \ee

We define Lie algebra valued gauge potentials with respect to this
decomposition,
\be A \ = \ A^k L_k + A^{\rho \ k}_{\ \ a} \, T_{\rho \ k}^{\ \ a}, \ee
which are differential forms in two dimensions
\be A \ = \ dz \, A_z(z,\bz) + d\bz \, A_\bz (z,\bz). \ee
The gauge transformations are given as
\be {}^g A \ = \ g A g^{-1} + gdg^{-1}, \ee
where the group elements $g$ depend on the parameters
$\al^-,\al^0,\al^+$ for the \slt subalgebra and
$\al^{\rho \ k}_{\ \ a}$ for the remaining generators.
The parameters are functions of $z$ and $\bz$.

We define, as usual, covariant derivative and field strength.
For $\Sigma (z,\bz)$ transforming as
\be {}^g \Sigma \ = \ g \Sigma, \ee
in some representation if the Lie group, the covariant derivative is
\be D \Sigma \ = \ d \Sigma + A \Sigma, \ee
with $A$ in the appropriate representation. We will, in the sequel,
frequently use the term {\em matter fields} for $\Sigma$. Applying the
covariant  exterior derivative once more one obtains
\be DD\Sigma \ = \ F \Sigma, \ee
with
\be F \ = \ dA - AA, \ee
the covariant field strength which satisfies Bianchi identities
\be dF \ = \ FA-AF, \ee
as a consequence of the nilpotency of the exterior derivative.

The BRS symmetry, arising originally in the quantization of gauge
theories, has an interpretation \cite{Sto77},\cite{Sto84}
as a differential algebra with an
additional grading, related to the appearance of the Lie algebra valued
Faddeev-Popov ghost fields \be \omega \ = \ \omega^k L_k + \omega^{\rho
\ k}_{\ \ a} \, T_{\rho \ k}^{\ \ a}. \ee Correspondingly a nilpotent  BRS
operation is defined as \bea
sA &=& -d \omega + A \omega + \omega A, \\
s\omega &=& \omega\omega,
\ena
and
\be s\Sigma \ = \ -\omega \Sigma. \ee
The BRS grading is often referred to as ghost number. While the exterior
derivative raises the form degree by one unit and leaves the ghost
number unchanged, the BRS operator raises the ghost number by one unit
and does not change the form degree. The nilpotency properties are
$d^2 = 0$, $s^2 = 0$ and $ds+sd = 0$.

Hence this system of fields and derivations describes a gauge structure
in a special basis of its Lie algebra in terms of a bigraded differential
algebra.

It is important to observe that the gauge potentials have
well defined conformal properties, $A_z$ and $A_\bz$ are
conformally covariant of weights $(1,0)$ and $(0,1)$, respectively.
Under a conformal change of coordinates
\be z \mapsto w(z), \cem \bz \mapsto \bw(\bz), \ee

they change as
\be A_w \ = \ \frac{1}{w'} A_z, \cem A_\bw \ = \ \frac{1}{\bw'} A_\bz.
\ee In this local description the internal gauge symmetry and conformal
transformations do not interfere with each other.

\indent

So far we have presented the standard gauge BRS structure mostly
in order to fix our notations. We shall now
describe a procedure which allows to identify for each \slt spin occuring
in the Lie algebra decomposition \begin{itemize}
\item a primary field $W^\rho_{a+2}$ of conformal weight $(a+2,0)$
      corresponding to the $\cw$-currents,
\item a conformally covariant field $v^{\rho \ -a-1}_\bz$ of conformal
      weight $(-a-1,1)$, corresponding to the $\cw$-gauge fields,
\item a ghost field $c^{\rho \ -a-1}$ of weight $(-a-1,0)$, conformally
      covariant as well, corresponding to $\cw$-gauge transformations
\end{itemize}
together with the complete nilpotent BRS algebra realized on this set
of fields and on those arising in the \slt substructure itself, which
are
\begin{itemize}
\item a projective connection $\Lzz$ which ensures conformal covariance
      due to its inhomogeneous transformation law,
\item a covariant $(-1,1)$ differential $\vz$, which was proposed
      to play the role of a Beltrami differential in
      refs.(\cite{BFK91},\cite{DHR90},\cite{GLM91}),
\item a ghost field $c^z$ of weight $(-1,0)$ which has a certain
      ressemblance with the diffeomorphism ghost arising in other
      contexts \cite{BB87,B88}.
\end{itemize}

\indent

The basic ingredients in this prescription are field dependent
redefinitions of the gauge potentials which have the form of gauge
transformations, highest weight pa\-ra\-me\-tri\-za\-tion,
 and zero curvature conditions. The \slt substructure plays an
important special role in the discussion of conformal covariance.

\indent

\subsect{Highest weight and conformal parametrization}

\indent

In a first step we consider a gauge transformation which depends on
the parameter $\al^0$ only,
\be g_0 \ = \ e^{\al^0 L_0}. \ee
Since $L_0$ measures the \slt spin component, all the gauge fields
transform covariantly according to
\be {}^{g_0}  A^{\rho \ k}_{\ \ a} \ = \  A^{\rho \ k}_{\ \ a}  \
e^{-k\al^0}, \ee except for $A^0$, the gauge potential pertaining to the
generator $L_0$, which picks up an inhomogeneous term:
\be ^{g_0} A^0 \ = \ A^0 - d \al^0. \ee
In particular, for
\be A^- \ = \ dz \, A_z^- + d\bz \, A_\bz^-, \ee
one has
\be {}^{g_0} A^- \ = \ A^- \, e^{\al^0}. \ee

We define now what we will call the {\em conformal parametrization}.
It consists in a redefinition on the set of gauge potentials which
has the form of a gauge  transformation, denoted $\hat{g}_0$ and chosen
such that ({\em cf}. also \cite{Pol90}).
\be {}^{\hat{g}_0} A_z^- \ = \ 1, \ee
\ie $\hat{\al}^0 \ = \ -\log A_z^-$.
In general, in the conformal parametrization, we define
\be \Ga \ = \ {}^{\hat{g}_0} A. \lbl{confgauge} \ee
This redefinition assigns now definit conformal weights to any of the
gauge potentials due to the transformation properties of $A_z^-$ and
the definition
\be \Ga^{\rho \ k}_{\ \ a} \ = \ A^{\rho \ k}_{\ \ a} \lp A_z^-\rp^k,  \ee
\ie $\Ga^{\rho \ k}_{\ \ a}$ has conformal weight $(k,0)$.

In the \slt substructure itself the conformal parametrization is
particularly relevant, there we define
\be
\Ga^- \ = \ dz + d\bz \, \frac{A_\bz^-}{A_z^-}
            \ \stackrel{\rm def}{=} \ dz + d\bz \, \vz \ = \ v^z,
\lbl{gam-1}
\ee
inducing a constant term which will be crucial in the subsequent
investigations.
At $k=0$ inhomogeneous derivative terms appear,
\bea
\Ga^0 &=& dz \, (A_z^0 + \dz \, \log A_z^-)
                 + d\bz \, (A_\bz^0 + \dbz \log A_z^-) \nn \\
      &\stackrel{\rm def}{=}& dz \, \cz + d\bz \, \cbz \ = \ \chi,
\lbl{gam-0}
\ena
and for $k=+1$ one obtains
\be
\Ga^+ \ = \ dz \, {A_z^+}{A_z^-}
               + d\bz \, {A_\bz^+}{A_z^-}
    \ \stackrel{\rm def}{=} \ dz \, \lzz + d\bz \, \lbzz \ = \ \lz.
\lbl{gam+1}
\ee
In these equations we have used suggestive index notations to account for
the {\em soldering} of internal \slt and conformal properties. By
construction, the new quantities appearing here are inert under $g^0$
gauge transformations. In exchange, they acquire well-defined conformal
properties. For instance,
$\vz$ is now a conformally covariant tensor of weight $(-1,1)$,
\be {v_\bw}^w \ = \ \vz \frac{w'}{\bw'}, \ee
$\lzz$ is a quadratic differential of weight $(2,0)$,
\be \lambda_{ww} \ = \ \frac{1}{(w')^2} \lzz, \ee
whereas $\cz$ transforms inhomogeneously according to
\be \chi_w \ = \ \frac{1}{w'}\lp \cz - \frac{w''}{w'} \rp. \ee
Due to this particular transformation law, $\cz$ will play the role of
a gauge potential - it will serve to define covariant derivatives with
respect to conformal transformations. All the remaining
gauge potential components are conformally
covariant, their weights are determined from the respective index
structure.

\indent

In analogy with the gauge potentials we define
\be c \ = \ \hat{g}_0 \om \hat{g}_0^{-1} + \hat{g}_0 s \hat{g}_0^{-1},
\lbl{coghost} \ee
the conformal parametrization for the ghost fields (remember the
gauge-like redefinition induced by $\hat{g}_0$ is field-dependent).
With this definition the BRS transformations in the conformal
parametrization take the form
\bea
s \, \Ga &=& -dc+\Ga c +c \Ga, \\
s \, c &=& c \, c.
\ena
The ghost fields are decomposed as
\be c \ = \ c^k L_k + c^{\rho \ k}_{\ \ a} \, T_{\rho \ k}^{\ \ a},
\lbl{confghost} \ee
with conformally covariant coefficients $c^k$ and $c^{\rho \ k}_{\ \ a}$
of weights $(k,0)$.

Similarly, for the fields $\Sigma$ we define the conformal
parametrization
\be \Psi \ = \ {}^{\hat{g}_0} \Sigma,
\lbl{confmat} \ee
with covariant derivative
\be \cd(\Ga) \Psi \ = \ d\Psi + \Ga \Psi \ = \ {}^{\hat{g}_0} D(A)
\Sigma, \ee and BRS transformation
\be s \, \Psi \ = \ - c \Psi. \ee

\indent

So far we have established the {\em conformal parametrization \/} for  the
gauge potentials, the ghost sector and the matter fields.
We come now  back to the discussion of the gauge potentials where
we impose, in a next step, the {\em highest weight gauge condition}
\be A_{z \ a}^{\rho \ k} \ = \ 0, \cem \mbox{for} \cem -a-1 \leq k \leq a, \ee
\ie all the $z$-components of the gauge potentials are constraint
to be zero, except for the highest weight components
at $k=a+1$, where we define
\be W_{a+2}^\rho \ = \ \Ga_{z \ a}^{\rho \ a+1}, \ee
indicating that this component is a conformal tensor of weight $(a+2,0)$.
As to the existence of this highest weight gauge and its implications
for residual gauge transformations we refer to \cite{DS84} and
\cite{FOR*92} and references quoted there.

For the $\bz$-components of the gauge potentials, on the other hand,
we define at lowest weight, $k=-a-1$, for each \slt spin,
\be v_\bz^{\rho \ -a-1} \ = \ \Ga_{\bz \ a}^{\rho \ -a-1}, \ee
whereas for the $\cw$-ghost fields at lowest weight we introduce
the notation
\be c^{\rho \ -a-1} \ = \ c^{\rho \ -a-1}_{\ a}. \ee
Recall that $v_\bz^{\rho \ -a-1}$ are conformal $(-a-1,1)$
differentials and the ghosts $c^{\rho \ -a-1}$ have conformal
weight$(-a-1,0)$. We also shall use the
convention $c^-=c^z$ and $c^+=c_z$ in the \slt subsector to emphasize
the conformal tensorial properties.

In the discussion at the end of this chapter we will see that
zero field strength conditions in the conformal
parametrization reduce the number of variables
considerably. At the end one is left with
$v_\bz^{\rho \ -a-1}$, $W_{a+2}^\rho$, $c^{\rho \ -a-1}$ for each
irreducible  representation and with $\vz$, $\cz$, $\lzz$ and $c^z$, $c_z$
for the
\slt substructure. All the other components of the gauge and ghost
fields will be recursively expressed in terms of these few basic
variables and their conformally covariant derivatives. Moreover, the BRS
algebra for these basic variables emerges and one derives expressions for
$\prt_\bz W_{a+2}^\rho$ in terms of conformally covariant operators
and differential polynomials, with similar remarks applying for the
sector of the covariantly constant fields $\Psi$.

\indent

\subsect{Projective parametrization}

\indent

Before turning to the detailed discussion of zero curvature conditions
we will now introduce the definition of what we call {\em projective
parametrization}.  This will again be a redefinition which has the form
of a gauge  transformation.
It will have the effect to eliminate $\cz$ and to introduce, at the same
time, a projective connection $\Lzz$, replacing the quadratic differential
$\lzz$. Likewise, the ghost $c_z$ disappears from the set of independent
variables. This goes as follows. We consider a gauge transformation which
depends on the parameter $\al^+$ only,
\be g_+ \ = \ e^{\al^+ L_+}. \ee
As a general property, this gauge transformation acts inside a given
representation of \slt spin. Since $L_+$ is the positive step operator,
$g_+$ acts always as a finite polynomial in $\al^+$ such that in the
transformation law of a given component $A_{\ \ a}^{\rho \ k}$ only
contributions of gauge potentials $A_{\ \ a}^{\rho \ k'}$ with
$k' \leq k$ occur. It follows that the lowest weight fields
$\vz$, $c^z$ and $v_\bz^{\rho \ -a-1}$, $c^{\rho \ -a-1}$
are inert under those transformations.
Moreover, in the highest weight gauge the fields $W_{a+2}^\rho$ are
invariant as well.

Let us first have a closer look to what happens in the \slt substructure
itself.
Since gauge transformations are defined on the original
fields $A_z, A_\bz$, the transformation laws of the
fields in the conformal parametrization are easily obtained
from direct substitution in the relevant definitions.
It is not hard to see that for $\cz$ and $\lzz$ one obtains
\bea
{}^{g^+} \cz &=& \cz + 2 \eta_z,
\\
{}^{g^+} \lzz &=& \lzz -\dz \eta_z + \cz \eta_z + \eta_z \eta_z,
\ena
observing that the gauge parameter $\alpha^{+1}$ appears always in the
combination \be \eta_z \ = \ \alpha^{+1} A_z^{-1}, \ee
assigning conformal dimension $(1,0)$ to the gauge parameter $\eta_z$
in the conformal pa\-ra\-me\-tri\-za\-tion.
We define now the {\em projective parametrization},
\be
\Pi \ = \ {}^{\hat{g}^+}\Ga.
\lbl{proj1} \ee
It is obtained from the conformal parametrization through a redefinition
which has the form of a gauge transformation
\be \hat{g}^+ \ = \ g(0,0,\hat{\alpha}^{+1}),
\lbl{proj2}\ee
such that
\be \hat{\eta}_z \ = \ \hat{\alpha}^{+1}  A_z^{-1} \ = \ - \shalf \cz.
\lbl{proj3}\ee

It is, of course, understood that the gauge transformations
are evaluated on the original variables and then substituted in the
conformal parametrization. Following this prescription one obtains
\bea
{}^{\hat{g}^+} \cz &=& 0, \\
{}^{\hat{g}^+} \lzz &=& \sm{1}{2} \Lzz \ = \
          \lzz + \sm{1}{2} \dz \cz - \sm{1}{4} \cz \cz.
\ena

As a consequence of the inhomogeneous transformations of $\cz$ under
conformal transformations one finds that the combination
\be \pi_{zz} \ = \ \dz \cz - \sm{1}{2} \cz \cz, \ee
transforms as a projective connection,
\be \pi_{ww} \ = \ \frac{1}{(w')^2} \lp \pi_{zz} - \{w,z\} \rp, \ee
with a Schwarzian derivative
\be
\{w,z\} \ = \ \lp \frac{w''}{w'} \rp^{'} - \frac{1}{2} \lp \frac{w''}{w'}
\rp^2,
\ee
as inhomogeneous term. Since $\lzz$ is a covariant quadratic differential,
the combination
\be \Lzz \ = \ 2\lzz + \pi_{zz}, \ee
transforms as a projective connection as well.

This discussion shows that, in the projective parametrization, $\cz$
disappears and $\lzz$ is replaced by the projective connection $\Lzz$.
Moreover,
as already remarked above, for each \slt spin, the lowest weight
fields $\vz$, $v_\bz^{\rho \ -a-1}$ and their ghosts
$c^z$ and  $c^{\rho \ -a-1}$ as well as
the highest weight fields $W_{a+2}^\rho$ do not change in the
transition from the conformal to the projective parametrization,
\ie they remain conformally covariant.
On the other hand, the remaining fields
$\cbz$, $\lbzz$ and $\Ga^{\rho \ k}_{\bz \ a}$ for $k \geq -a$, as well
as the corresponding ghosts are no longer conformally covariant when
expressed in the conformal parametrization due to the appearance
of $\cz$ in the redefinitions. But these fields
will be eliminated recursively by means of the zero curvature conditions
as will be discussed below.

\indent

\subsect{Zero curvature conditions}

\indent

We turn now to the detailed discussion of the {\em zero curvature
conditions}. On the one hand we shall argue in terms of the conformal
parametrization, where conformal covariance is manifest due to the
presence of
$\cz$ which behaves as a gauge potential under conformal transformation
and appears in the recursive procedure such that at each step
successively covariant derivatives emerge. This can be seen quite
clearly in the explicit expressions for the field strength two-forms:  in
the conformal parametrization they read
\bea
F^- &=& (d-\chi)v^z
          -\shalf \Ga^{\rho \ k}_{\ \ a} \ \Ga^{\sigma \ l}_{\ \ b} \
              \si_{\rho k \ \sigma l}^{\ \, a \ \ \, b \ -}, \\
F^0 &=& d\chi +2v^z \lz
          -\shalf \Ga^{\rho \ k}_{\ \ a} \ \Ga^{\sigma \ l}_{\ \ b} \
              \si_{\rho k \ \sigma l}^{\ \, a \ \ \, b \ 0}, \\
F^+ &=& (d+\chi)\lz
          -\shalf \Ga^{\rho \ k}_{\ \ a} \ \Ga^{\sigma \ l}_{\ \ b} \
              \si_{\rho k \ \sigma l}^{\ \, a \ \ \, b \ +},
\ena
for the \slt substructure
whereas for the remaining generators one obtains
\bea
F^{\tau \ m}_{\ \ c} &=& (d + m \chi) \Ga^{\tau \ m}_{\ \ c}
              -v^z \, \Ga^{\tau \ m+1}_{\ \ c} \ \si_{\tau \ m+1}^{- \ c}
\nonumber \\
             & & -\lz \, \Ga^{\tau \ m-1}_{\ \ c} \ \si_{\tau \ m-1}^{+ \
c}
              -\shalf \Ga^{\rho \ k}_{\ \ a} \ \Ga^{\sigma \ l}_{\ \ b} \
              \si_{\rho k \ \sigma l \ \ \ c}^{\ \, a \ \ \, b \ \tau m},
\ena

We note first of all the appearance of conformally covariant derivatives
$d+m\chi$. Moreover, we observe that the quadratic terms
involving $v^z$ actually have a linear piece due to the constant term
in the definition of $v^z=dz+d\bz \, \vz$. For
a given $a$, these linear terms occur in all components
$-a-1 \leq k \leq a$ of the field strengths
except in the highest weight ones ($k=a+1$).
For vanishing field strength this means
then that the coefficients $\cbz$, $\lbzz$ and $\Ga^{\rho \ k}_{\bz \ a}$
for $k
\geq -a$ are  recursively expressed
in terms of the basic covariant variables $\vz$, $\lzz$ of the
\slt subsector and $v^{\rho \ -a-1}_\bz$, $W^\rho_{a+2}$ of
the various spins occuring in the decomposition. Conformal
covariance is ensured at each step through the covariant derivatives
involving $\cz$.
The zero field strength conditions
at the highest weights give then rise to equations which express
$\prt_\bz \lzz$ and  $\prt_\bz W^\rho_{a+2}$ in terms of those basic
variables and their conformally
covariant derivatives. In particular, for each $a$ a differential
operator of order $2a+3$ occurs which maps conformal weight $-a-1$ into
conformal weight $a+2$, and which is of the form
\be (\prt_z + (a+1) \cz) \cdots (\prt_z - (a+1) \cz) \, v^{\rho \
-a-1}_\bz. \ee But there are also other contributions which have the form
of differential polynomials in the basic variables. The explicit form of
these  ''anholomorphicity equations'' depends of course on the structure
of the Lie algebra and the particular decomposition under consideration.

\indent

What happens in the ghost sector? To discuss this issue consider
the explicit form of the BRS transformations, which, in the \slt
sector are given as
\bea
s \, v^z &=& -(d-\chi) \, c^z +v^z c^0
          +\Ga^{\rho \ k}_{\ \ a} \ c^{\sigma \ l}_{\ \ b} \
              \si_{\rho k \ \sigma l}^{\ \, a \ \ \, b \ -}, \\
s \, \chi &=& - dc + 2 \lz c^z - 2v^z c_z
          + \Ga^{\rho \ k}_{\ \ a} \ c^{\sigma \ l}_{\ \ b} \
              \si_{\rho k \ \sigma l}^{\ \, a \ \ \, b \ 0}, \\
s \, \lz &=& -(d+\chi) \, c_z + \lz c^0
          + \Ga^{\rho \ k}_{\ \ a} \ c^{\sigma \ l}_{\ \ b} \
              \si_{\rho k \ \sigma l}^{\ \, a \ \ \, b \ +},
\ena
and for the remaining generators take the form
\bea
s \, \Ga^{\tau \ m}_{\ \ c}
&=& -(d + m \chi) \, c^{\tau \ m}_{\ \ c} -m \, c^0 \,  \Ga^{\tau \ m}_{\
\ c} + \lp v^z \, c^{\tau \ m+1}_{\ \ c} + c^z \, \Ga^{\tau \ m+1}_{\ \ c}
\rp
               \si_{\tau \ m+1}^{- \ c} \nonumber \\
& & + \lp \lz \, c^{\tau \ m-1}_{\ \ c} + c_z \, \Ga^{\tau \ m-1}_{\ \ c}
\rp
                  \si_{\tau \ m-1}^{+ \ c}
             + \Ga^{\rho \ k}_{\ \ a} \ c^{\sigma \ l}_{\ \ b} \
              \si_{\rho k \ \sigma l \ \ \ c}^{\ \, a \ \ \, b \ \tau m}.
\ena
Here, a similar mechanism as before takes place. Note first of
all, that these equations are differential one forms of BRS
grade one. As a consequence each of these equations has a
component in the direction $dz$ and another one in the direction $d\bz$
both of ghost number one.

Let us first discuss the \slt subsector. In its $dz$ component,
the first of the three equations determines $c^0$ as a dependent
variable, the second equation shows that the BRS transformation
of $\cz$ has a term linear in $c_z$,
\be s \, \cz \ = \ -2 c_z + \cdots, \ee
and the third one determines $s \, \lzz$. In the $d\bz$ sector
the first equation yields $s \, \vz$, while the two other ones
contain no new information.
For the remaining generators, in the $dz$ components,
the constant term allows to
express the ghost fields $c^{\rho \ k}_{\ \ a}$, for $k \geq -a$,
in terms of the basic fields. The highest weight
equations in $dz$ determine then the BRS transformation of
$W^\rho_{a+2}$. Here again, among other things, the differential operator
of order $2a+3$ shows up, acting on $c^{\rho \ k}_{\ \ a}$.
On the other hand, the $d\bz$
components of these equations determine, at lowest weight,
the BRS transformations of $v^{\rho \ -a-1}_\bz$,
whereas all the other $d\bz$ equations, \ie for $k \geq -a$
should be identically satisfied with the information extracted
so far.

It remains to discuss the equations at BRS grade two. In the \slt
subsector they read
\bea
s \, c^z &=& c^0 \, c^z
          + \shalf c^{\rho \ k}_{\ \ a} \ c^{\sigma \ l}_{\ \ b} \
              \si_{\rho k \ \sigma l}^{\ \, a \ \ \, b \ -}, \\
s \, c^0 &=& 2c_z \, c^z
          + \shalf c^{\rho \ k}_{\ \ a} \ c^{\sigma \ l}_{\ \ b} \
              \si_{\rho k \ \sigma l}^{\ \, a \ \ \, b \ 0}, \\
s \, c_z &=& c_z \, c^0
          + \shalf c^{\rho \ k}_{\ \ a} \ c^{\sigma \ l}_{\ \ b} \
              \si_{\rho k \ \sigma l}^{\ \, a \ \ \, b \ +},
\ena
where the first and the third equation determine the BRS operation
on $c^z$ and $c_z$, respectively, while the second equation contains
no new information. Likewise, for the remaining ghosts,
\bea
s \, c^{\tau \ m}_{\ \ c}
&=&  c^z \, c^{\tau \ m+1}_{\ \ c} \si_{\tau \ m+1}^{- \ c}
-m \, c^0 \, c^{\tau \ m}_{\ \ c} \nonumber \\
& & + c_z \, c^{\tau \ m-1}_{\ \ c} \si_{\tau \ m-1}^{+ \ c}
+ \shalf c^{\rho \ k}_{\ \ a} \ c^{\sigma \ l}_{\ \ b} \
\si_{\rho k \ \sigma l \ \ \ c}^{\ \, a \ \ \, b \ \tau m},
\ena
at each $a$, the lowest weight equations
determine the BRS transformations of the independent ghost fields
$c^{\rho \ -a-1}$, and all the remaining equations
are identically satisfied.

This completes the discussion of zero curvature conditions and the
construction of a nilpotent BRS differential algebra in terms of
the conformal parametrization and the highest weight gauge. The
presentation was deliberately rather qualitative with the intention
to explain rather the general structure than detailed quantitative
features. Those can be most conveniently studied in explicit examples,
which will be given in detail later on in this paper for the
cases of \slt and \sltr.

One of the most important features on which we would like to insist
here, however, is the conformal covariance of the whole construction,
as a consequence of the presence of conformally covariant derivatives
in terms of $\cz$.

In the transition from the conformal to the projective parametrization
by means of the $\hat{g}^+$ redefinition, on the other hand, $\cz$
(as well as $c_z$, the corresponding ghost) disappear and the basic
field and ghost variables left over after the recursive procedure of
the zero curvature conditions are
\be \vz, \cem c^z, \cem \Lzz, \ee
in the \slt subsector and
\be v^{\rho \ -a-1}_\bz, \cem c^{\rho \ -a-1}, \cem W^\rho_{a+2}, \ee
for each \slt spin occuring in the decomposition. All the fields
are conformally covariant except $\Lzz$, the projective connection.

What happens to conformal covariance in the projective parametrization?
The answer is that conformal covariance is maintained in terms of
the projective connection. This is due to the fact that the transition
from the conformal to the projective parametrization consists in
redefinitions which have the form of gauge transformations and leave
therefore invariant the zeros in the curvature conditions.

As a consequence, the conformal covariance of
the differential operators and differential polynomials occuring in the
anholomorphicity relation for projective connection $\Lzz$ and for the
$\cw$-currents $W^\rho_{a+2}$ and in the BRS differential algebra is
achieved solely in terms of $\Lzz$.

In particular, one recovers, for each value of $a$, a
covariant differential operator $\Delta^{2a+3}$ which maps
$(-a-1,k)$-differentials into $(a+2,k)$-differentials \cite{FIZ91},  \ie
\be \Delta^{2a+3} : \cem \cv^{-a-1} \ \mapsto \ \cv^{a+2}. \ee

So much for the general discussion, we shall turn now to
more detailed descriptions of specific examples.
\indent

\sect{$SL(2)$ GAUGE STRUCTURE: THE CORNER-STONE}

\indent

As we have pointed out in the preceding general discussion,
the Lie algebra \slt plays a crucial role in the
construction of $\cw$-gauge structures: a given Lie algebra can
give rise to various different $\cw$-gauge structures
according to different \slt embeddings.
The particular structure of the \slt embedding is
responsible for the special properties of a soldering of
internal and conformal symmetries,
especially the appearance of higher conformal spins.

It is therefore worthwile to study first the case of \slt itself
in some detail. We shall consider here \slt-valued gauge potentials
together with a doublet field transforming covariantly
under gauge transformations.
On this basic set of fields we will then rediscuss
in full detail the properties of
the {\em conformal} and the {\em projective} parametrizations,
in particular the appearance of the {\em projective connection},
and the explicit form of the residual gauge transformations \cite{Gar93}.

The {\em zero curvature conditions}  express the
anholomorphicity of the projective connection in terms of a
conformally covariant differential operator, known also from
the second Hamiltonian structure of the KdV hierarchy.

We consider then the zero curvature condition as
{\em integrability condition} for a covariantly constant
doublet field. The condition of vanishing covariant derivative
determines one of the doublet fields in terms of the other one,
which is then subject to the (conformally covariant)
Sturm-Liouville equation.

Finally we present the nilpotent {\em differential BRS algebra},
which strongly ressembles with the factorized
diffeomorphism BRS algebra encountered in the context of
two dimensional conformal field theory and construct algebraically
the corresponding consistent anomaly.

\indent

\subsect{Conformal and projective parametrizations }

\indent

We start from a gauge potential one-form
\be
A \ =\ A^k L_k,
\ee
which takes its values in the Lie algebra of \slt
with generators $L_k, \, k=-1,0,+1$ and commutation relations
\be [L_k,L_l] \ = \ (k-l) L_{k+l}. \ee
Correspondingly the gauge transformations depend on three
parameters $\alpha^k(z,\bz)$ and we denote an element of
the gauge group $g(\alpha^k)=g(\alpha^-,\alpha^0,\alpha^+)$.

In addition, we consider a doublet with respect to these \slt
gauge transformations, which we denote
\be
\Sigma \ =\ \lp \begin{array}{c} \Sigma_{+1/2} \\ \Sigma_{-1/2}
\end{array} \rp.
\ee
In this representation the generators are taken to be
\be
\lambda_- \ = \ \lp \begin{array}{cc} 0&0\\ -1&0 \end{array} \rp, \cem
\lambda_{0} \ = \ \lp \begin{array}
          {cc} -\sm{1}{2}&0 \\ 0&+\sm{1}{2} \end{array} \rp, \cem
\lambda_+ \ = \ \lp \begin{array}{cc} 0&+1 \\ 0&0 \end{array} \rp.
\ee
The covariant exterior derivative is defined as
\be D(A) \, \Sigma \ = \ (d+A^k \lambda_k) \Sigma, \ee
with the usual definition of field strength
\be F(A) \ = \ dA - AA. \ee
This is our basic set of classical fields.

\indent

In a first step, we consider now a special gauge transformation
$g^0=g(0,\alpha^0,0)=\exp(\alpha^0 L_0)$. On the gauge potentials
themselves this gives rise to
\bea
{}^{g^0}A^- &=& A^- e^{+\alpha^0}, \nn \\
{}^{g^0}A^0 &=& A^0 - d \, \alpha^0, \lbl{confparam} \\
{}^{g^0}A^+ &=& A^+ e^{-\alpha^0}, \nn
\ena
whereas the doublet transforms as
\be {}^{g^0}\Sigma_{+1/2} \ = \ e^{-\alpha^0/2} \Sigma_{+1/2}, \cem
    {}^{g^0}\Sigma_{-1/2} \ = \ e^{+\alpha^0/2} \Sigma_{-1/2}. \ee
Taking into account the one-form nature of the gauge potentials,
we consider in particular
\be
{}^{g^0}A^- \ = \ dz A_z^- \, e^{+\alpha^0}
                   + d\bz A_\bz^- \, e^{+\alpha^0}, \ee

In view of these equations we will now perform a
particular redefinition of
the gauge potential components and of the doublet field,
which has the form of such a gauge transformation,
\ie for non-vanishing $A_z^-$ we define
the {\em conformal parametrization}:
\be
\Ga \ = \ {}^{\hat{g}^0}A, \cem \Psi \ = \ {}^{\hat{g}^0} \Sigma,
\ee
of field dependent parameter such that
\be
{\hat{g}^0} \ = \ g(0,\hat{\alpha}^0,0),  \cem
          \hat{\alpha}^0 \ = \ - \log A_z^-.
\ee

\indent

Let us look at these redefinitions in some more detail.
For the gauge potential components at $k=-1$,
\be
\Ga^- \ = \ dz + d\bz \, \frac{A_\bz^-}{A_z^-}
            \ \stackrel{\rm def}{=} \ dz + d\bz \, \vz \ = \ v^z,
\ee
a constant term is induced.
At $k=0$, inhomogeneous derivative terms appear,
\bea
\Ga^0 &=& dz \, (A_z^0 + \dz \, \log A_z^-)
                 + d\bz \, (A_\bz^0 + \dbz \log A_z^-) \nn \\
      &\stackrel{\rm def}{=}& dz \, \cz + d\bz \, \cbz \ = \ \chi,
\ena
and for $k=+1$ one obtains
\be
\Ga^+ \ = \ dz \, {A_z^+}{A_z^-}
               + d\bz \, {A_\bz^+}{A_z^-}
    \ \stackrel{\rm def}{=} \ dz \, \lzz + d\bz \, \lbzz \ = \ \lz.
\ee

In these equations we have used suggestive notations to account for the
soldering of internal \slt and conformal properties. By construction, the
new quantities appearing here are inert under $g^0$ gauge transformations.
In exchange, they acquire well-defined conformal properties. For instance,
$\vz$ is now a conformally covariant tensor of weight $(-1,1)$,
\be {v_\bw}^w \ = \ \vz \frac{w'}{\bw'}, \ee
$\lzz$ is a quadratic differential of weight$(2,0)$,
\be \lambda_{ww} \ = \ \frac{1}{(w')^2} \lzz, \ee
whereas $\cz$ transforms inhomogeneously,
\be \chi_w \ = \ \frac{1}{w'}\lp \cz - \frac{w''}{w'} \rp.
\lbl{chitra} \ee
Due to this particular transformation law, $\cz$ will play the role of
a gauge potential, it will serve to define covariant derivatives with
respect to conformal transformations. All the remaining fields are
conformally
covariant, their weights are determined from the respective index
structure.

A similar soldering occurs for the matter fields, where we define
\be
\Psi_{+1/2}\ =\ \sqrt{A_z^-} \, \Sigma_{+1/2}
                \ \stackrel{\rm def}{=} \ \psi_\zt, \cem
\Psi_{-1/2} \ =\ \frac{1}{\sqrt{A_z^-}} \, \Sigma_{-1/2}
                \ \stackrel{\rm def}{=} \ \psi^\zt.
\ee
Here greek indices are used to indicate
the occurence of half-integer conformal dimensions,
\be
\psi_\om \ = \ \frac{1}{\sqrt{w'}} \, \psi_\zt, \cem
\psi^\om \ = \ {\sqrt{w'}} \, \psi^\zt,
\ee
in other words, $\psi_\zt$ is a $(\shalf,0)$ and $\psi^\zt$ a
$(-\shalf,0)$-differential.

Since the redefinitions used to define the conformal parametrization
have the form of gauge transformations we define field strength and
covariant derivative in the conformal parametrization as
\be
\cf(\Ga) \ = \ {}^{\hat{g}^0} F(A), \cem
             \cd(\Ga) \Psi \ = \ {}^{\hat{g}^0} D(A) \Sigma.
\ee
In some more detail, this yields
\bea
\cf^- &=& d v^z + v^z \, \chi, \\
\cf^0 &=& d \chi + 2 v^z \, \lz, \\
\cf^+ &=& d \lz  - \lz \, \chi,
\ena
for the field strength and
\bea
\cd \psi_\zt &=& (d-\sm{1}{2}\chi)\psi_\zt  + \lz \psi^\zt, \\
\cd \psi^\zt &=& (d+\sm{1}{2}\chi)\psi^\zt  - v^z \psi_\zt,
\ena
for the covariant derivatives.

\indent

In a second step we consider transformations which depend on
the parameter $\alpha^+$ only,
$g^+=g(0,0,\alpha^+)=\exp(\alpha^+ L_+)$,
defined on the original fields $A_z, A_\bz$ and $\Sigma$.
The transformation laws of the
fields in the conformal parametrization are obtained
from substitution in the relevant definitions.
It is not hard to see that for $v^z$, $\chi$ and $\lz$ one obtains
\bea
{}^{g^+} v^z &=& v^z, \nn\\
{}^{g^+} \chi &=& \chi + 2 \eta_z v^z, \lbl{g+} \\
{}^{g^+} \lz &=& \lz - d \eta_z + \eta_z \chi + \eta_z \eta_z v^z. \nn
\ena
The gauge parameter $\alpha^+$ appears always in the combination
\be \eta_z \ =\ \alpha^+ A_z^-, \ee
it acquires conformal dimension $(1,0)$.
Observe that $v^z$ is invariant under these gauge transformation,
and the same holds for $\psi^\zt$.

The {\em projective parametrization}, defined as
\be
\Pi \ = \ {}^{\hat{g}^+}\Ga, \cem \Psi_\pi \ = \ {}^{\hat{g}^+}\Psi,
\ee
is obtained from the conformal parametrization through a redefinition
which has the form of a gauge transformation
\be \hat{g}^+ \ = \ g(0,0,\hat{\alpha}^+), \ee
such that
\be \hat{\eta}_z \ = \ \hat{\alpha}^+  A_z^- \ = \ - \shalf \cz. \ee
Using the explicit form of the $g^+$ gauge transformations given above
one finds easily from the transformation laws \rf{g+}
\bea
{}^{\hat{g}^+} \vz &=& \vz, \\
{}^{\hat{g}^+} \cz &=& 0, \\
{}^{\hat{g}^+} \lzz &=&  \lzz + \sm{1}{2} \dz \cz - \sm{1}{4} \cz \cz
           \stackrel{\rm def}{=}\sm{1}{2} \Lzz.
\ena

Clearly, the new variable $\Lzz$ will transform inhomogeneously
under conformal transformations. This is due to the appearance
of the combination
\be \pi_{zz} \ = \ \dz \cz - \sm{1}{2} \cz \cz. \ee
Taking into account eq.\rf{chitra} for the transformation of $\cz$
one recognizes easily that $\pi_{zz}$
behaves in the same way as a projective connection, \ie
\be \pi_{ww} \ = \ \frac{1}{(w')^2} \lp \pi_{zz} - \{w,z\} \rp, \ee
with a Schwarzian derivative
\be
\{w,z\} \ = \ \lp \frac{w''}{w'} \rp^{'}
                - \frac{1}{2} \lp \frac{w''}{w'} \rp^2,
\ee
as inhomogeneous term.
Since $\lzz$ is a covariant quadratic differential, the combination
\be \Lzz \ = \ 2\lzz + \pi_{zz}, \ee
transforms as a projective connection as well.

\indent

So far we made use of the $g^0$ and $g^+$ gauge transformations to
reduce the \slt gauge structure to the conformal and
the projective parametrizations.
Recall that the fields in the conformal parametrization
are inert under $g^0$ transformations while the
fields in the projective parametrization are invariant
under both, $g^0$ and $g^+$ transformations. What about the
remaining residual gauge transformations of parameter $\alpha^-$ ?

Performing a gauge transformation
$g^- = g(\alpha^-,0,0) = \exp(\alpha^-L_-)$
on the original variables and keeping track of the various redefinitions
one learns that $\alpha^-$ always occurs in the combination
\be
\eta^z \ = \ \frac{\alpha^-}{A_z^-},
\ee
{\em \ie} it acquires conformal dimension $(-1,0)$ and thus looks like
a vector field. In terms of this
parameter the corresponding residual transformations read then
\bea
{}^{g^{-}}\vz &=& \frac{1}{\Omega}\,\left(\,
\vz\,-\,\left(\,\partial_{\bz}\,+\,\cbz\,\right)\,\eta^z\,+\,
\lbzz\,\eta^z\,\eta^z\,\right) \\
{}^{g^{-}} \lzz &=& \lzz\,\Omega \label{g-}\\
{}^{g^{-}}\cz &=& \cz\,-\,2\lzz\,\eta^z\,+\,\frac{1}{\Omega}\,\dz\Omega
\ena
with
\be
\Omega \ = \ 1\,-\,\left(\,\dz\,+\,\cz\,\right)\,\eta^z\,
+\,\lzz\,\eta^z\,\eta^z,
\ee
It is instructive to consider the infinitesimal version of these
transformations. For $\vz$ one obtains
\be \delta_{g^-} : \ \ \ \vz \ \mapsto \
\vz\,-\, \dbz \eta^z + \vz \, \dz\, \eta^z
\,-\, \eta^z \, \left( \, \cbz \,-\, \vz \cz\,\right), \ee
whereas the projective connection $\Lzz$ transforms as
\be \delta_{g^-} : \ \ \ \Lzz \ \mapsto \
\Lzz - \eta^z \dz \Lzz -2 \Lzz \dz \eta^z - \dz \dz \dz \eta^z. \ee

For the doublet fields we obtain
for a finite transformation
\bea
{}^{g^{-}} \psi^\zeta &=& \frac{1}{\sqrt{\Omega}}\,
		(\psi^\zeta - \eta^z \psi_\zeta), \\
{}^{g^{-}} \psi_\zeta &=& \sqrt{\Omega} \psi_\zeta ,
\ena
while for the infinitesimal version
\bea
\delta_{g^-} : \ \ \ \psi^\zeta & \mapsto & \psi^\zeta
		+ \shalf \psi^\zeta ( \dz + \cz )\,\eta^z
		-  \eta^z\, \psi_\zeta , \\
\delta_{g^-} : \ \ \ \psi_\zeta & \mapsto & \psi_\zeta
		-  \shalf \psi_\zeta ( \dz + \cz )\,\eta^z .
\ena
Although these are special \slt transformations we observe certain
similarities with diffeomorphism transformations of parameter
$\eta^z$. This will become even more striking after having taken into
account the zero curvature conditions in the next section (they will
have the effect to replace $\cbz - \vz \cz \ $ by $\ \dz \vz$).

\indent

\subsect{Zero curvature conditions}

\indent

We consider now the doublet field $\Sigma$ to be covariantly constant,
\ie
$D(A) \, \Sigma = 0$. Applying covariant exterior derivative to this
condition implies vanishing field strength, $F(A)=0$. Since the transition
to the conformal parametrization has the form of a gauge transformation,
these conditions are invariant, and we shall investigate them here
in the conformal parametrization:
\be
\cd(\Ga) \Psi \ = \ 0, \cem  \cf(\Ga) \ = \ 0.
\ee
In some more detail, for the doublet $\Psi$, the conditions
of zero covariant derivative read
\bea
(d-\sm{1}{2}\chi)\psi_\zt  + \lz \psi^\zt &=& 0, \\
(d+\sm{1}{2}\chi)\psi^\zt  - v^z \psi_\zt &=& 0,
\ena
whereas the zero curvature conditions for the gauge potentials are
\bea
d v^z + v^z \chi &=& 0, \\
d\chi + 2 v^z \lz &=& 0, \\
d\lz - \lz \chi &=& 0.
\ena
Since $v^z$ contains a constant term, three of the above equations
allow to express certain fields in terms of others and their derivatives.
For the doublet this yields
\be
\psi_\zt \ = \ \left(\,\dz\,+\,\sm{1}{2}\,\cz\,\right)\,\psi^\zt,
\lbl{doublet}
\ee
while for the gauge potentials,
the coefficients $\cbz$ and $\lbzz$ can be expressed as
\bea
\cbz &=& \lp \dz + \cz \rp \vz, \\
\lbzz &=& \vz \lzz + \sm{1}{2} \dz \lp \dz + \cz \rp \vz
           -\sm{1}{2} \dbz \cz.
\ena
Observe that all the derivatives occuring here are conformally
covariant thanks to the inhomogeneous transformation law of $\cz$.

Substitution of these expressions in the other equations gives then
rise to differential expressions involving the remaining basic fields
$\vz$, $\cz$, $\lzz$ and $\psi^\zt$. As to the latter, the zero
covariant derivative conditions yield
\be
\dbz \,\psi^\zt \ = \ \vz\,\dz\,\psi^\zt - \sm{1}{2}\,\psi^\zt\,\dz\,\vz,
\ee
and
\be
\left(\,\dz\,-\,\sm{1}{2}\,\cz\,\right)\,
\left(\,\dz\,+\,\sm{1}{2}\,\cz\,\right)\,\psi^\zt\,+\,\lzz\,\psi^\zt \  =
\  0.
\ee
Again, the differential operator appearing here is conformally covariant.
Straightforward manipulation shows that it can be rewritten in terms of
$\pi_{zz}$:
$$
\left(\,\dz\,-\,\sm{1}{2}\,\cz\,\right)\,
\left(\,\dz\,+\,\sm{1}{2}\,\cz\,\right) \ = \
\dz\dz\,+\,\sm{1}{2}\,\pi_{zz},
$$
and one obtains the conformally covariant Sturm-Liouville equation
\be
\left(\,\dz\dz\,+\,\sm{1}{2}\,\Lzz\,\right)\,\psi^\zt \ = \ 0,
\ee
in terms of the second-order differential operator
\be \2 \ = \ \dz\dz\,+\,\sm{1}{2}\,\Lzz. \ee
Absorbing $\cz$ in a redefinition of $\lzz$ amounts to pass from the
conformal to the projective parametrization, note also that this
particular redefinition has the form of a Miura transformation
\cite{Mat90}.
The covariance of the resulting equation is ensured by construction.
The differential operator $\2$ provides a map from the space $\cv^{-1/2}$
of covariant $(-\shalf,k)$-differentials into $\cv^{+3/2}$, the space
of covariant $(\sm{3}{2},k)$-differentials:
\be \2 : \cem \cv^{-1/2} \ \mapsto \ \cv^{+3/2}. \ee

On the other hand, in the zero curvature conditions,
substituting for $\cbz$ and $\lbzz$
in the third equation yields, with very
little algebraic effort, the conformally covariant equation
\bea
\lefteqn{ \dbz \lp \lzz + \sm{1}{2} \dz \cz - \sm{1}{4} \cz \cz \rp \ = }
\nonumber \\[2mm]
& & \cem 2 \lzz \dz \vz + \vz \dz \lzz
    + \sm{1}{2} \lp \dz - \cz \rp \dz \lp \dz + \cz \rp \vz.
\ena
Again, a straightforward reshuffling in the third order differential
operator gives rise to
\be
\lp \dz - \cz \rp \dz \lp \dz + \cz \rp \vz \ = \
\lp \dz \dz \dz + \dz \cdot \pi_{zz} + \pi_{zz} \dz \rp \vz,
\ee
absorbing again $\cz$ in the same redefinition as already encountered
above,
reflecting the transition from the conformal to the projective
parametrization
by means of a Miura transformation of $\lzz$.
As a result we are simply left with
\be
\dbz \Lzz \ = \ \lp \dz \dz \dz + \dz \cdot \Lzz  + \Lzz \dz \rp \vz,
\ee
where the third order differential operator
\be
\3 \ = \ \dz \dz \dz  + \dz \cdot \Lzz + \Lzz \dz,
\ee
maps covariant $(-1,k)$-differentials into covariant
$(2,k)$-differentials:
\be \3 : \cem \cv^{-1} \ \mapsto \ \cv^{+2}. \ee

Observe that the same differential operator appears also in
the second hamiltonian structure of the KdV equation. These
intriguing structures gave rise to investigations concerning
possible relations between the anomalous conservation
equation of the energy momentum tensor in two dimensional conformal
theory and the KdV hierarchy \cite{BFK91}, \cite{DHR90}.

As already anticipated above, after taking into account the zero
curvature conditions, the
infinitesimal $\eta^z$ transformations of $\vz$, $\Lzz,$
and $\psi^\zeta$ acquire a form which is very similar to diffeomorphisms.
This is in particular the case for $\vz$ which
transforms under $\eta^z$ transformations in the same way
as the Beltrami differential transforms under changes of coordinates,
\ie
\be \delta_{g^-} : \ \ \ \vz \ \mapsto \
\vz\,-\, \dbz \eta^z + \vz \, \dz\, \eta^z
\,-\, \eta^z \, \dbz \vz, \lbl{resv} \ee
the transformation of the projective connection has already been given
above,
\be \delta_{g^-} : \ \ \ \Lzz \ \mapsto \
\Lzz - \eta^z \dz \Lzz -2 \Lzz \dz \eta^z - \dz \dz \dz \eta^z.
\lbl{resL} \ee
Finally, using \rf{doublet}, we obtain for the doublet field
\be
\delta_{g^-} : \ \ \ \psi^\zeta \ \mapsto \
\psi^\zeta - \eta^z \dz \psi^\zeta + \shalf \psi^\zeta \dz \eta^z.
\lbl{resmatt} \ee

The analogues of these gauge structures for other Lie algebras are
sometimes called $\cw$-diffeo\-morphisms, in the \slt case one might
call them $\cw_2$-diffeomorphisms. A very convenient way to treat
these structures is in terms of BRS differential algebra, the subject
of the next section.

\indent

\subsect{The BRS differential algebra and anomaly structure}

\indent

As is well known, the BRS differential algebra of gauge and
ghost fields has a compact formulation in terms of the
generalized objects
\be \tilde{A} \ = \ A + \omega, \ee
which unifies gauge and ghost fields, and the generalized
nilpotent derivation
\be \tilde{d} \ =\ d + s, \ee
unifying exterior derivative and the BRS operation. These
generalized quantities are then used to define a generalized
field strength
\be \tilde{F}(\tilde{A}) \ = \ \tilde{d} \tilde{A} - \tilde{A}
\tilde{A}, \ee and a generalized covariant derivative
\be \tilde{D}(\tilde{A}) \, \Sigma \ = \ ( \tilde{d} + \tilde{A}) \,
\Sigma. \ee The BRS transformations of gauge and ghost fields are then
recovered from the horizontality conditions \cite{Sto84}
\be \tilde{F}(\tilde{A}) \ = \ F(A), \ee
which, for the matter fields take the form
\be \tilde{D}(\tilde{A}) \, \Sigma \ = \ D(A) \, \Sigma. \ee

To be definite,
we shall work in the conformal parametrization defined as
\be
\tilde{\Gamma} \ = \ {}^{\hat{g}^0} \tilde{A},
\ee
and, in accordance with previous notations, we define
\be
\tilde{v}^z \ = \ v^z + c^z, \cem
\tilde{\chi} \ = \ \chi + c, \cem
\tilde{\lambda}_z \ = \ \lz + c_z.
\lbl{ghost}
\ee
Also, we shall work right away with the zero curvature conditions,
such that the horizontality equations become
\bea
\tilde{d} \tilde{v}^z + \tilde{v}^z \tilde{\chi} &=& 0, \\
\tilde{d} \tilde{\chi} + 2 \tilde{v}^z \tilde{\lambda}_z &=& 0, \\
\tilde{d} \tilde{\lambda}_z - \tilde{\lambda}_z \tilde{\chi} &=& 0.
\ena
Going through this set of equations at ghost number one allows,
first of all, to express the dependent variables
\bea
c &=& (\dz + \chi_z) c^z, \\
\lbl{BRSsup}
c_z \ + \shalf s \chi_z &=& \shalf \dz c + c^z \lzz.
\ena
Given this information, the remaining equations reduce then
to a BRS differential algebra, with $s^2=0$, which closes on
the basic variables $\vz$, $c^z$ and
$\Lzz$ in the following simple way (see also \cite{Zuc93a}):
\bea
s \vz &=& \dbz c^z - \vz \dz c^z + c^z \dz \vz, \\
s c^z &=& -c^z \dz c^z, \\
s \Lzz &=& \lp \dz \dz \dz  + \dz \cdot \Lzz  + \Lzz \dz \rp c^z.
\ena
This differential algebra reproduces exactly the infinitesimal action,
see eqs.\rf{resv} and \rf{resL}, of the residual transformations.

Similarly, for the covariantly constant doublet field in the
conformal parametrization one obtains
\bea
(\tld{d}-\sm{1}{2}\tld{\chi})\psi_\zt + \tld{\lambda}_z \psi^\zt &=& 0,
\\ (\tld{d}+\sm{1}{2}\tld{\chi})\psi^\zt - \tld{v}^z \psi_\zt &=& 0.
\ena
Using the relations derived so far, in particular
\rf{doublet} and \rf{BRSsup}, this set of equations
reduces to the BRS transformation of $\psi^\zt$,
\be
s \psi^\zt \ =\ c^z \dz \psi^\zt - \shalf \psi^\zt \dz c^z
\ee
clearly exhibiting the conformal nature of the field $\psi^\zt$ as a
$(-\shalf,0)$-differential ({\em cf}. eq.\rf{resmatt}).

This concludes our discussion of the differential BRS algebra of
gauge and ghost fields in presence of the zero curvature conditions and
of covariantly constant matter fields. We will use the notion of
$\cw_2$-gauge structure for the set of fields
\be \vz, \cem \Lzz, \cem c^z, \cem \psi^\zt, \ee
subject to the BRS transformations just derived and to the equations
\be
\dbz \Lzz \ = \ \lp \dz \dz \dz + \dz \cdot \Lzz  + \Lzz \dz \rp \vz,
\ee
and
\be
\dbz \,\psi^\zt \ = \ \vz\,\dz\,\psi^\zt - \sm{1}{2}\,\psi^\zt\,\dz\,\vz,
\ee
together with
\footnote{in fact, formally this equation might be read as
$\Lambda \ = \ \prt \Sigma - \frac{1}{2} \Sigma \Sigma$,
with $\Sigma \ = \ -2 \frac{1}{\psi} \, \prt \, \psi$}
\be
\left(\,\dz\dz\,+\,\sm{1}{2}\,\Lzz\,\right)\,\psi^\zt \ = \ 0,
\ee
arising from the conditions of vanishing curvature and
covariant derivative.

\indent

We come now to the discussion of possible anomalies as solutions
of the consistency conditions. That is one asks for a local
functional $\ca_{\bz z}^{(1)}$, which should be a $(1,1)$
differential of ghost number one, constructed in terms of
the set of basic fields, in our present case $\vz$, $\Lzz$
and $c^z$, and
which is closed under BRS transformations up to total derivatives, \ie
\be s \ca_{\bz z}^{(1)} \ = \ \dz \ca_\bz^{(2)} + \dbz \ca_z^{(2)}. \ee
It is rather easy to see that the expression
\be \ca_{\bz z} \ =
    \ c^z \dbz \Lzz - \vz s \Lzz \ = \ c^z \3 \vz - \vz \3 c^z, \ee
provides indeed a solution to the consistency conditions
and an explicit computation allows to identify
\bea
\ca_z^{(2)} &=& c^z \3 c^z, \\
\ca_\bz^{(2)} &=& \left( c^z \dz \vz - \vz \dz c^z \right) \dz \dz c^z
                    - c^z \dz c^z \, \dz \dz \vz.
\ena
We wish to emphasize that $\ca_{\bz z}^{(1)}$ as well as $\ca_z^{(2)}$
and $\ca_\bz^{(2)}$ are conformally covariant tensors.
Also, the formal similarity with the factorized conformal anomaly
(in terms of a Beltrami differential and a background holomorphic
projective connection) appears quite clearly \cite{Laz90,KLT90}.

Solving the consistency condition is not enough to characterize
an anomaly, it must also be nontrivial: one has to convince
oneself that it is not possible to express the solution
given here as the BRS variation of a local functional in
the basic fields $\vz$, $\Lzz$ and $c^z$ up to derivative terms.
More explicitly one has to show that
\be \ca_{\bz z}^{(1)} \neq s {\cal B}_{\bz z}^{(0)}
   + \dbz {\cal B}_z^{(1)} + \dz {\cal B}_\bz^{(1)}.
\ee
This can indeed be confirmed by explicit inspection of possible
counterterms, taking into account the restrictions arising
from the index structures ({\em viz.} conformal weights)
and the polynomial form and degrees of derivatives in the
expression for the anomaly.

On the other hand, as to the BRS transformations
of the ghost number two partners of the anomaly,
an explicit computation shows that
\bea
s \ca_z^{(2)} &=& \dz \ca^{(3)}, \\
s \ca_\bz^{(2)} &=& \dbz \ca^{(3)},
\ena
with
\be \ca^{(3)} \ = \ c^z \dz c^z \, \dz \dz c^z, \ee
which is a conformally invariant tensor of ghost number three.
This sequence of $s$ modulo $d$ equations can be compactly
summarized in terms of descend equations, again in striking analogy
with the usual conformal anomaly (ch. III.2 in  \cite{Laz90}).
To this end we define
\be
\tld{\anom} \ = \ \anom_2^1 + \anom_1^2 + \anom_0^3,
\ee
with obvious reference of the indices to form degree and BRS grading
(ghost number) and identify
\bea
\anom_2^1 &=& - \shalf\, dz \wedge d\bz \ \ca_{\bz z}^{(1)}, \\
\anom_1^2 &=& \shalf dz \, \ca_z^{(2)} - \shalf d\bz \, \ca_\bz^{(2)}, \\
\anom_0^3 &=& \shalf \ca^{(3)}.
\ena
In this notation (the factors of one half are for later convenience)
we obtain the descend equations in the form
\be
s\, \anom^1_2 + d\, \anom^2_1 \ =\ 0, \cem
s\, \anom^2_1 + d\, \anom^3_0 \ =\ 0, \cem
s\, \anom^3_0 \ =\ 0,
\lbl{desct}
\ee
or, even more compactly,
\be \tld{d}\, \tld{\anom} \ =\ 0. \ee

\indent

In our approach, $\cw$-gauge structures are obtained from
the usual BRS gauge structure of \slt valued Yang-Mills gauge
potentials and their ghosts via zero curvature conditions
in combination with the conformal, and, finally, projective
parametrization. In the following we would like to point out
that the anomaly obtained above in terms of the basic variables of
the projective parametrization fits into this picture as well.
It can indeed be related to the usual
construction of the Yang-Mills anomaly via descend equations
\cite{Sto84,MSZ85}. Recall that there a (nontrivial) solution
to the consistency conditions is identified as the ghost number
one component of the generalized Chern-Simons form, constructed
from $\tld{A} = \tld{A}^k \lambda_k$ and $\tld{d}$, \ie the
generalized three form
\be
\tld{\cqb}(\tld{A}) \ =\ {\rm tr} \lp \tld{A}\, \tld{d} \tld{A}
			- \sm{2}{3} \tld{A} \tld{A} \tld{A}\rp ,
\ee
with contributions
\be
\tld{\cqb} \ =\ \cqb^1_2 + \cqb^2_1 + \cqb^3_0,
\ee
at various levels of ghost number (upper index) and
form degree (lower index). The consistent anomaly is then
identified in $\cqb^1_2$.

Keeping in mind that the transition from the original \slt gauge
structure to the conformal parametrizations has the form of a gauge
transformation, $\tld{\Ga} \ =\ {}^{\hat{g}^0}\tld{A}$, one has
\be
\tld{\cqb}(\tld{A}) \ =\ \tld{\cqb}(\tld{\Ga}) + \tld{d}\, \tld{\Theta},
\ee
due to the well-known fact (the so-called triangular equation
\cite{MSZ85})  that the Chern-Simons form changes by the exterior
derivative of a two form under gauge transformations.
In our particular case it is straightforward to obtain explicitly
\be \tld{\Theta} \ =\
\sm{1}{2} \, \tld{\chi} \ \frac{1}{A_z^-} \, \tld{d}\, A_z^- . \ee
Moreover, since we are dealing here with flat gauge potentials, the
Chern-Simons three form reduces to
\be \tld{\cqb}(\tld{\Ga}) \ =\
      \sm{1}{3}\, {\rm tr} \lp \tld{\Ga} \tld{d}\, \tld{\Ga} \rp
\ =\ \sm{1}{3}\, {\rm tr} \lp \tld{\Ga} \tld{\Ga} \tld{\Ga} \rp,
\ee
and it is easy to convince oneself that, in the conformal parametrization
({\em cf.} also \cite{BaBeGr89}, \cite{BBG91}),
\be \tld{\cqb}(\tld{\Ga}) \ =\ \tld{v}^z  \tld{\chi}\, \tld{\lambda}_z.
\ee

Finally, an explicit calculation shows that
\be \tld{v}^z  \tld{\chi}\, \tld{\lambda}_z \ = \
     \tld{\anom} + \tld{d}\, \tld{\Xi}, \ee
where the trivial contributions
\be \tld{\Xi} \ = \ \Xi_2^0 + \Xi_1^1 + \Xi_0^2, \ee
are given in terms of the variables of the conformal parametrization
as follows:
\bea
\Xi_2^0 &=& - \shalf dz \wedge d\bz \left(
      \dz \dz \vz + \cz \dz \vz + 2 \vz \Lzz \right), \\
\Xi_1^1 &=& - \shalf dz \left(
      \dz \dz c^z + \cz \dz c^z + 2 c^z \Lzz \right)
      + \shalf d\bz \, \cz \left( c^z \dz \vz - \vz \dz c^z \right), \\
\Xi_0^2 &=& -\shalf \cz \, c^z \dz c^z.
\ena
In fact, this decomposition amounts to the explicit transition from
the conformal to the projective parametrization (which, as should be  kept
in mind, is provided by a field redefinition which has the form
of a particular gauge transformation on the original \slt variables).

To summarize, we have established the explicit relation
\be \tld{\cqb}(\tld{A}) \ = \
\sm{1}{3}{\rm tr} \left( \tld{A} \tld{A} \tld{A} \right)
\ = \ \tld{\anom} + \tld{d} \left( \tld{\Theta} + \tld{\Xi} \right).  \ee
By definition, as a generalized differential three-form,
$\tld{\cqb}(\tld{A})$ is a conformally invariant quantity. On the other
hand, as a result of our explicit construction, $\tld{\anom}$ is
invariant under conformal transformations as well. As a consequence, the
counterterms
\be \tld{U} \ = \ \tld{\Xi} + \tld{\Theta}, \ee
should change under a conformal transformation as
\be \tld{U}(w) \ = \ \tld{U}(z) + \tld{d} \tld{u}, \ee
with $\tld{u} = u_1^0 +u_0^1$. Indeed, an explicit calculation
yields
\bea
u_1^0 &=& -\shalf \left( dz \, \log A_z^- + d \bz \, \vz \right)
               \frac{w''}{w'}, \\
u_0^1 &=& -\shalf c^z \, \frac{w''}{w'}.
\ena

\indent

\sect{\sltr AND $\cw_3$-GAUGE STRUCTURES}

\indent

We shall discuss now the case of \sltr as one of the simplest
examples which nevertheless illustrates already the most important
features relevant to the construction in the general case.

First of all, and as is well known, \sltr allows for two
different \slt decompositions.
The first one, which gives rise to what is usually called $\cw_3^{(1)}$,
is related to the principal \slt embedding, in this case the eight
generators of \sltr are split into three plus five (\ie  \slt spin two).

The $\cw_3^{(2)}$-gauge structure, on the other hand, is based on a
decomposition where the splitting is $3+2+2+1$, in other words two spin
1/2 doublets and a singlet in addition to the \slt generators, this
decomposition is quite familiar in elementary particle physics since
the days of the eight-fold way.

We shall separately examine these two possibilities in full
detail, along the lines of the general discussion
presented in the beginning of this paper.

Particular emphasis will be on the construction
of the complete nilpotent BRS algebra, which may be
understood to represent the infinitesimal $\cw$-transformations
and their commutators. Moreover, this differential algebra
will be used to determine explicitly anomalies, both for $\cw_3^{(1)}$
and for $\cw_3^{(2)}$, as solutions of the consistency conditions.

Finally, the properties of covariantly constant matter fields
will be presented in detail.

The presentation will proceed in two parallel tracks for the
two different embeddings.
In both cases we shall, after some
motivating remarks, present the results right away in the projective
parametrization.

\indent

\subsect{$\cw_3^{(1)}$-gauge structure}

\indent

In the language of the general dicussion of chapter 2, the \sltr Lie
algebra is here represented in terms of a set of
three generators $L_k,\, k=-1,0,+1$, for the \slt
subalgebra and another set of five generators $T_{\rho \ k}^{\ \ a}$ with
$a = 1$ and $m = -2,-1,0,+1,+2$, representing spin two with respect to
the \slt subalgebra.
Since this representation of dimension five occurs just once, the label
$\rho$  can be neglected and we denote these generators simply $T_m$
(omitting the index $a=1$ as well). The commutation
relations in  terms of this decomposition are then given as
\bea
\left[L_k,L_l\right] &=& (k-l) \, L_{k+l}, \\
\left[L_k,T_m \right] &=& (2k-m) \, T_{m+k}, \\
\left[T_m,T_n\right] &=& - \sm{1}{3} \, (m-n) \, (2m^2 + 2n^2 - mn -8)
\, L_{m+n}.
\ena
Following the general procedure we define the gauge potential one-form
and the ghost fields in this decomposition as
\bea
A &=& A^k L_k + A_1^m T_m, \\
\om &=& \om^k L_k + \om_1^m T_m,
\ena
and introduce covariant matter fields as a triplet of \sltr,
\be
\Sigma \ = \ \left(
\begin{array}{c}
\Sigma_+ \\ \Sigma_0 \\ \Sigma_-
\end{array}
\right).
\ee
In this three-dimensional representation
we use the following $3 \times 3$ matrices for the generators:
\cbe
\begin{array}{l}
L_-\ =\ \left( \begin{array}{ccc}
0 & 0 & 0 \\ - \sqrt{2} & 0 & 0 \\ 0 & - \sqrt{2} & 0
\end{array} \right),\ \
L_0\ =\ \left( \begin{array}{ccc}
-1 & 0 & 0 \\ 0 & 0 & 0 \\ 0 & 0 & 1
\end{array} \right), \ \
L_+\ =\ \left( \begin{array}{ccc}
0 & \sqrt{2} & 0 \\ 0 & 0 & \sqrt{2} \\ 0 & 0 & 0
\end{array} \right), \\[1.4cm]
T_{-2}\ =\ \left( \begin{array}{ccc}
0 & 0 & 0 \\ 0 & 0 & 0 \\ -4 & 0 & 0
\end{array} \right), \ \
T_{-1}\ =\ \left( \begin{array}{ccc}
0 & 0 & 0 \\ - \sqrt{2} & 0 & 0 \\ 0 & \sqrt{2} & 0
\end{array} \right),\ \
T_0\ =\ \left( \begin{array}{ccc}
- \sm{2}{3} & 0 & 0 \\ 0 & \sm{4}{3} & 0 \\ 0 & 0 & - \sm{2}{3}
\end{array} \right),
\end{array}
\cee
\be
\begin{array}{lll}
T_{+1}\ =\ \left( \begin{array}{ccc}
0 & \sqrt{2} & 0 \\ 0 & 0 & - \sqrt{2} \\ 0 & 0 & 0
\end{array} \right), &
\ \ T_{+2}\ =\ \left( \begin{array}{ccc}
0 & 0 & -4 \\ 0 & 0 & 0 \\ 0 & 0 & 0
\end{array} \right). &
\end{array}
\lbl{repmat}
\ee
The explicit form of the matrices $L_k$ shows that $\Sigma$ is indeed
a triplet with respect to the \slt subalgebra (for notational simplicity
we use the same symbols for the generators and their specific matrix
realization).

\indent

As outlined in the general discussion, the {\em conformal parametrization}
is obtained from this set of gauge, ghost and and matter fields
by means of redefinitions, \rf{confgauge}, \rf{confghost},
\rf{confmat}, which have the form of a $\alpha^0$ gauge transformation
with the parameter identified as
$$\hat{\alpha}^0 \ =\ - \log A_z^-.$$
As we have seen, these redefinitions assign well defined conformal
properties.
Recall that the gauge potentials in the \slt subsector,
cf. \rf{gam-1}, \rf{gam-0} and \rf{gam+1}, become
\bea
v^z &=& dz + d\bz \, \vz, \\
\chi &=& dz \, \cz + d\bz \, \cbz, \\
\lz &=& dz \, \lzz + d\bz \, \lbzz.
\ena
They play a particularly important r\^ole.
For the remaining five components of the \sltr gauge potential
in the conformal parametrization the definitions are
\be
\Ga_1^m \ =\ {}^{\hat{g}^0} A_1^m \ =\ A_1^m (A_z^-)^m
\ =\ dz \, \Ga_{z \ 1}^{\ \ m} + d\bz \, \Ga_{\bz \ 1}^{\ \ m},
\lbl{conf2}
\ee
assigning conformal weights $(m,0)$ to $\Ga_1^m$ and, as a consequence,
conformal weights $(m+1,0)$ to $\Ga_{z \ 1}^{\ \ m}$ and $(m,1)$
to $\Ga_{\bz \ 1}^{\ \ m}$.

As a next step we impose the {\em highest weight constraints},
\be \Ga_{z \ 1}^{\ \ m} \ = \ 0, \cem m \ = \ -2, -1, 0, +1, \ee
and we denote
\be \Ga_{z \ 1}^{\,  +2} \ \stackrel{\rm def}{=} \ W_3, \ee
the remaining nonzero component, which has conformal weight $(3,0)$.
The $\bz$ components remain arbitrary, for the moment.However, the
relevant quantity here (which will survive after the zero-curvature
conditions) is
\be \Ga_{\bz \ 1}^{\, -2} \ \stackrel{\rm def}{=} \ \vzz, \ee
a conformal tensor of weight $(-2,1)$.

\indent

The {\em projective parametrization} is defined as a redefinition
of the gauge potentials which has the form of a $\alpha^+$ gauge
transformation (see eqs. \rf{proj1} ff). In the \slt subsector
it has the effect to eliminate $\cz$ at the expense of the appearance
of the projective connection $\Lzz$. In other words, the
projective parametrization establishes explicitly
the highest weight gauge in the \slt subsector.
It is important to note that the fields $\vz$, $\vzz$ and
$W_3$ remain unchanged in the transition from the conformal
to the projective parametrization (the latter, $W_3$, due to the
highest weight constraints), they remain covariant conformal tensors.
The other components,
$\Ga_{\bz \ 1}^{\ \ m}$, for $m \geq -1$,
will receive additional
contributions in terms of $\cz$ and its derivatives and therefore
acquire non-covariant conformal transformations. However, those
quantities will recursively disappear once the
{\em zero curvature conditions} are imposed.
The mechanism of this recursion procedure has been explained in
the general discussion of chapter 2, for the case at hand
we have to consider the three equations
\bea
\cf^- &=& d v^z + v^z \, \chi, \\
\cf^0 &=& d \chi + 2 v^z \, \lz + 16 \, \Ga_1^{+2} \, \Ga_1^{-2}, \\
\cf^+ &=& d \lz  - \lz \, \chi + 4 \, \Ga_1^{+2} \, \Ga_1^{-1},
\ena
for the \slt part (note the appearance of the additional terms
with $\Ga_1^m$ relative to the pure \slt case) and five equations
\be
\cf^m_1\ =\ d \Ga_1^m + m \chi\, \Ga_1^m
      + (m+3) v^z\, \Ga_1^{m+1} + (m-3) \lz\, \Ga_1^{m-1}
\ee
for the spin two sector, corresponding to the values $m = -2,-1,0,+1,+2$.
In the zero curvature conditions two of the first three equations are
recursion relations while the third one, after substitution gives rise
to the equation
\be
\dbz \Lzz \ = \ \3 \vz - 8 \left(2 \vzz \, \dz W_3 + 3 W_3 \, \dz \vzz
\right).
\lbl{hol1} \ee
On the other hand, in the second set the zero curvature conditions for
the values
$m=-2,-1,0,+1$ are recursive and at $m=+2$ one obtains
\be
\dbz W_3 \ = \ \sm{1}{24} \5 \vzz + \vz \, \dz W_3 + 3 W_3 \, \dz \vz.
\lbl{hol2} \ee
The third order differential operator
\be
\Delta^{(3)} \ = \ \prt^3 + \prt \Lambda + 2 \Lambda \prt,
\ee
appeared already in the pure \slt case, it provides a covariant
map from the space of $(-1,k)$ into $(2,k)$-differentials:
\be \3 : \cem \cv^{-1} \ \mapsto \ \cv^{+2}, \ee
while the fifth order differential operator
\be
\Delta^{(5)} \ = \ \prt^5
                   + 2 \prt^3 \Lambda + 10 \Lambda \prt^3
                   + 15 \prt \Lambda \prt^2
                   + 9 \prt^2 \Lambda \prt
                   + 16 \Lambda \prt \Lambda
                   + 16 \Lambda \Lambda \prt.
\ee
defines a covariant mapping
\be \Delta^{(5)} : \cem \cv^{-2} \ \mapsto \ \cv^{+3}. \ee
So far we have brushed over the properties of the \sltr gauge
potentials in the principal decomposition, subject to
conformal and projective parametrizations, highest weight constraints
and zero curvature conditions. The final result being that
the remaining basic degrees of freedom
\begin{center} $\vz, \cem \Lzz, \cem$ and $\cem \vzz, \cem W_3,$
\end{center}
are subject to the modified holomorphicity equations \rf{hol1}, \rf{hol2}.

\indent

We shall now go through the same discussion, {\em mutatis mutandis},
for the matter triplet $\Sigma$. Following the general discussion
we define the conformal parametrization
\be \Psi \ = \ {}^{\hat{g}_0} \Sigma, \ee
as in \rf{confmat}, which in the present case of a triplet field
gives rise to
\be
\Psi_+ \ =\ A_z^-\, \Sigma_+\ \stackrel{\rm def}{=}\ \psi_z, \cem
\Psi_0 \ =\ \Sigma_0 \ \stackrel{\rm def}{=} \ \psi, \cem
\Psi_- \ =\ \frac{1}{A_z^-}\, \Sigma_- \ \stackrel{\rm def}{=} \ \psi^z,
\ee
introducing notations which clearly exhibit the corresponding
conformal weights:
As the transition to the conformal parametrization has the form of
a gauge transformation, the covariant derivative is given as
\be \cd(\Ga) \Psi \ = \ {}^{\hat{g}^0} D(A) \Sigma. \ee
In more explicit terms and taking into account the particular
matrix representation \rf{repmat} given above this reads
\bea
\cd \psi_z &=& \left(d-\chi\right)\psi_z + \sqrt{2} \left(\lz+\Ga_1^{+1}
\right)\psi  - 4 \Ga_1^{+2} \, \psi^z, \\
\cd \psi &=& \left(d + \sm{4}{3} \Ga_1^{+1}\right)\psi - \sqrt{2}
\left(v^z  + \Ga_1^{-1}\right)\psi_z
		+ \sqrt{2}\left(\lz + \Ga_1^{+1}\right) \psi^z, \\
\cd \psi^z &=& \left(d+\chi\right)\psi^z - \sm{2}{3} \Ga_1^0 \, \psi^z
		- \sqrt{2}\left(v^z - \Ga_1^{-1}\right)\psi - 4 v^{zz} \psi_z.
\ena
Inspection of these equations for vanishing
covariant derivatives shows easily that again, due
to the presence of the terms proportional to $v^z$, the last two
equations serve to recursively eliminate the components $\psi$
and $\psi_z$ as functions of the other variables and their derivatives.
In the transition to the projective parametrization the component
$\psi^z$ does not change.
As final result of this procedure one is left with two equations,
due to the $dz$ and $d\bz$ components of the covariant derivative,
and which are
\be
\left( \3 - 8 W_3 \right) \psi^z \ =\ 0,
\ee
and
\bea
\dbz \psi^z &=& \vz\, \dz\,\psi^z - \psi^z \dz\, \vz + 2 \vzz \dz \dz
\psi^z - \dz \vzz \dz \psi^z \nn\\[2mm]
&&+\ \sm{1}{3} \psi^z \dz \dz \vzz
+ \sm{8}{3} \vzz\, \Lzz\,\psi^z.
\ena
Note that the last four terms in this equation are conformally covariant
by virtue of the projective connection.

\indent

We come now to the discussion of the ghost sector and the corresponding
differential BRS algebra. The conformal parametrization, defined in
the general case in \rf{coghost} and \rf{confghost}, reads in our case as
\be c \ = \ c^k L_k + c_1^m T_m, \ee
where the ghost fields $c^k$ and $c_1^m$ have conformal weights
$(k,0)$ and $(m,0)$, respectively. Following the prescriptions
of the general case, it is straightforward to convince oneself
that after the transition to the projective parametrization
and in presence of the highest weight constraints only
the conformally covariant ghost fields
\be c^- \ = \ c^z, \cem c_1^{-2} \ = \ c^{zz}, \ee
of conformal weights $(-1,0)$ and $(-2,0)$ survive, all the
other ones are recursively eliminated. Taking into account all
the properties of the gauge and ghost fields, the complete
nilpotent BRS differential algebra of the basic variables is given as
\bea
s \vz &=& \dbz c^z + c^z \dz \vz - \vz \dz c^z
		+ \sm{2}{3} \left(\vzz \dz^3 c^{zz} - c^{zz} \dz^3 \vzz \right) \nn\\
		&& + \dz c^{zz} \dz^2 \vzz - \dz \vzz \dz^2 c^{zz}
  + \sm{16}{3} \Lzz \left(\vzz \dz c^{zz} - c^{zz} \dz \vzz \right)
\\[2mm] s \vzz &=& \dbz c^{zz} + c^z \dz \vzz - \vz \dz c^{zz}
			+ 2 \left( c^{zz} \dz \vz - \vzz \dz c^z \right) \\[2mm]
s c^z &=& -c^z \dz c^z - \dz c^{zz} \dz^2 c^{zz}
		+ \sm{2}{3} c^{zz} \dz^3 c^{zz} + \sm{16}{3} \Lzz c^{zz} \dz c^{zz}
\\[2mm] s c^{zz} &=& - c^z \dz c^{zz} - 2 c^{zz} \dz c^z \\[2mm]
s \Lzz &=& \3 c^z - 8 \left(2 c^{zz} \dz W_3 + 3 W_3 \dz c^{zz}\right)
\\[2mm] s W_3 &=& \sm{1}{24} \5 c^{zz} + c^z \dz W_3 + 3 W_3 \dz c^z
\ena
Likewise, the BRS variation for the matter field comes out to be
\be
s \psi^z \ =\ c^z \dz \psi^z - \psi^z \dz c^z
	+ 2 c^{zz} \dz^2 \psi^z - \dz c^{zz} \dz \psi^z
	+ \sm{1}{3} \psi^z \dz^2 c^{zz} + \sm{8}{3} \Lzz \psi^z c^{zz}.
\ee
This BRS differential algebra reflects exactly the transformations
and their commutators obtained in \cite{OSSvN92} for induced
$\cw_3$-gravity.

\indent

Given the complete nilpotent BRS algebra we may ask for a solution
of the consistency condition, \ie the existence of a local functional
of the basic variables of ghost number one which is BRS closed
modulo exterior derivative. Such a quantity can indeed be constructed,
and it is given as
\be
\ca_{\bz z}^{(1)} \ =\ \left(c^z \dbz - \vz s\right) \Lzz
-8 \left(c^{zz} \dbz - \vzz s \right) W_3.
\ee
In more explicit terms
\bea
\ca_{\bz z}^{(1)}
&=& c^z \3 \vz - \vz \3 c^z - \sm{1}{3}
			\left(c^{zz} \5 \vzz - \vzz \5 c^{zz}\right) \nn\\
&& -\ 8\left(c^z \vzz - \vz c^{zz}\right) \dz W_3 \nn\\
&& -\ 24 W_3 \left(c^z \dz \vzz - \vzz \dz c^z
+ c^{zz} \dz \vz - \vz \dz c^{zz} \right),
\ena
where the leading terms coincide indeed with the expression obtained from
the study of induced $\cw_3$-gravity in \cite{OSSvN92}. One should also
keep in mind that the individual contributions containing
$c^z$ or $c^{zz}$  only are
not separately solutions to the consistency conditions, only the
particular combination given here is. This is a remnant of the
Chern-Simons  origin of the anomaly, \ie the anomaly can be obtained, via
conformal and  projective parametrisation plus highest weight constraint
from
\be \mbox{tr} \left. \left( \tld{A} \tld{d} \tld{A} \right) \right|_2^1
\ee up to trivial terms, as we have checked by an explicit calculation
for the leading terms taking into account the particular decomposition
defined in the beginning of this subsection.

\indent

\subsect{${\cal W}_3^{(2)}$-gauge structure}

\indent

While the principal \slt decompositions are given solely in terms of
integer gradings, other \slt embeddings allow for half-integer gradings
as well. This is the case for the second \sltr decomposition which
will be discussed here and wich will lead us to define the
${\cal W}_3^{(2)}$-gauge structure. Again, first of all,
three \slt generators $L_k,\, k=-1,0,+1$ are identified.
Note that, although we use the same
symbols as in the previous case, it should be kept in mind that
this \slt is identified in a different manner among the generators
\sltr.
The remaining generators $\, T_{\rho \ k}^{\ \ a} \, $
are now arranged in two doublets of $a=-1/2$ with $k=\pm 1/2$,
distinguished by the hypercharge index $\rho=-1,+1$ and a singlet,
which in our notation has $a=-1$, $k=0$, the hypercharge itself.
The generators in the two doublets will be denoted $\, T_{\rho \ k} \,$,
neglecting the index $a$, whereas the hypercharge will be identified
as $T_{\ 0}^{-1} = Y$.
The commutation relations in this basis are then given as
\bea
\left[L_k,L_l\right] &=&  (k-l) \,  L_{k+l}, \\
\left[L_k,T_{\rho \ m}\right] &=& (\shalf k -m) \, T_{\rho \ m+k}, \\
\left[Y,L_k\right] &=& 0, \\
\left[Y,T_{\rho \ k}\right] &=& - \rho \, T_{\rho \ k}, \\
\left[T_{\rho \ k},T_{\sigma \ l}\right]
&=& - \shalf(\rho - \sigma) \left( (k+l) + \shalf (\rho - \sigma) (k-l)
\right) L_{k+l} \nn \\ && - \sm{3}{4} (\rho - \sigma) (k-l)^2 \,  Y.
\ena

Gauge fields and ghost fields are decomposed along this basis as follows:
\bea
A &=& A^k L_k + A^{\rho \; m} \, T_{\rho \; m} + A_Y Y, \\
\om &=& \om^k L_k + \om^{\rho\ m} \, T_{\rho \ m} + \om_Y Y,
\ena
with summation over $\rho=-1,+1$ and $m=-1/2,+1/2$. As before
we introduce a triplet of matter fields,
\be
\Sigma \ = \ \left(
\begin{array}{c}
\Sigma_{+1/2} \\ \Sigma_{-1/2} \\ \Sigma_0
\end{array}
\right),
\ee
but now the three-dimensional representation of \sltr is given  explicitly
in terms of the matrices
\be
\begin{array}{lll}
L_-\ =\ \left( \begin{array}{ccc}
0 & 0 & 0 \\ -1 & 0 & 0 \\ 0 & 0 & 0
\end{array} \right), &
L_0\ =\ \left( \begin{array}{ccc}
- \shalf & 0 & 0 \\ 0 & \shalf & 0 \\ 0 & 0 & 0
\end{array} \right), &
L_+\ =\ \left( \begin{array}{ccc}
0 & 1 & 0 \\ 0 & 0 & 0 \\ 0 & 0 & 0
\end{array} \right), \\[1cm]
\begin{array}{l}
T_{\ominus \ +1/2} \ =\ \left( \begin{array}{ccc}
0 & 0 & 1 \\ 0 & 0 & 0 \\ 0 & 0 & 0
\end{array} \right), \\[1cm]
T_{\oplus \ +1/2} \ =\ \left( \begin{array}{ccc}
0 & 0 & 0 \\ 0 & 0 & 0 \\ 0 & 1 & 0
\end{array} \right),
\end{array} &
\begin{array}{l}
T_{\ominus \ -1/2} \ =\ \left( \begin{array}{ccc}
0 & 0 & 0 \\ 0 & 0 & 1 \\ 0 & 0 & 0
\end{array} \right), \\[1cm]
T_{\oplus \ -1/2} \ =\ \left( \begin{array}{ccc}
0 & 0 & 0 \\ 0\  & 0 & 0 \\ 1 & 0 & 0
\end{array} \right),
\end{array} &
Y \ =\ \left( \begin{array}{ccc}
\sm{1}{3} & 0 & 0 \\ 0 & \sm{1}{3} & 0 \\ 0 & 0 & - \sm{2}{3}
\end{array} \right)
\end{array}
\lbl{matr2} \ee
Again the explicit form of the matrices $L_k$ justifies the notational
conventions in the components of $\Sigma$,
it decomposes into a doublet and a singlet
with respect to the \slt substructure. We denote negative and positive
hypercharges by the symbols $\ominus$ and $\oplus$, respectively,
in order to distinguish the hypercharge from the other indices.

\indent

What about the {\em conformal parametrization} in this case. We know
from the general discussion, that it is always defined in the same
way, \ie as a redefinition which has the form of a $\alpha^0$
gauge transformation of parameter $\hat{\alpha}^0 = - \log A_z^-.$
The details depend, however, on the special basis chosen for the
generators of the Lie algebra. In the present case the appearance
of half-integer $a$- and $k$-values in the Lie algebra
decomposition will give rise to bosonic
degrees of freedom of half-integer conformal weights among the
gauge, ghost and matter fields. More explicitly, the gauge
potentials at $a=-1/2$, $m=\pm 1/2$ in the conformal parametrization are
defined as
\be
\Ga^{\rho \ m} \ =\ {}^{\hat{g}^0} A^{\rho \ m} \ =\ A^{\rho \ m}
(A_z^-)^m,
\ee
whereas for $a=-1$ we have
\be
\Ga_Y \ =\ {}^{\hat{g}^0} A_Y \ =\  A_Y.
\ee
At $a=0$ we employ, of course, always the same definitions in terms
of $\vz$, $\chi$ and $\lz$.
For this \slt subsector the curvature in the conformal parametrization
reads
\bea
\cf^- &=& d v^z + v^z \, \chi
+ \Ga^{\ominus \ -1/2} \ \Ga^{\oplus \ -1/2}, \\
\cf^0 &=& d \chi + 2 v^z \, \lz
+ \Ga^{\ominus \ +1/2} \ \Ga^{\oplus \ -1/2}
				- \Ga^{\ominus \ -1/2} \ \Ga^{\oplus \ +1/2}, \\
\cf^+ &=& d \lz  - \lz \, \chi
- \Ga^{\ominus \ +1/2} \ \Ga^{\oplus \ +1/2}.
\ena
Recall that we use the symbols $\ominus$ and $\oplus$
to label quantities of hypercharge $\rho=-1$ and $\rho=+1$,
respectively. In the singlet sector $a=-1$ and $m=0$ the
field strength components are then given as
\be
\cf_Y \ =\ d \Ga_Y
- \sm{3}{2} \Ga^{\ominus \ -1/2} \ \Ga^{\oplus \ +1/2}
- \sm{3}{2} \Ga^{\ominus \ +1/2} \ \Ga^{\oplus \ -1/2},
\ee
and for the doublet at $a=-\shalf$ one obtains
\bea
\cf^{\rho \, -1/2} &=& d \Ga^{\rho \, -1/2}
	- \shalf \chi\, \Ga^{\rho \, -1/2}
 - \rho \, v^z\, \Ga^{\rho \, +1/2}
	+ \rho \, \Ga_Y \ \Ga^{\rho \, -1/2}, \\
\cf^{\rho \, +1/2} &=& d \Ga^{\rho \, +1/2}
	+ \shalf \chi\, \Ga^{\rho \, +1/2}
+ \rho \, \lz\, \Ga^{\rho \, -1/2}
	+ \rho \, \Ga_Y \ \Ga^{\rho \, +1/2}.
\ena
The highest weight constraints in this decomposition are simply
\be \Ga_z^{\rho \ -1/2} \ = \ 0. \ee
The independent fields are then identified as follows
\bea
\Ga_Y &=& dz\, W_1 + d\bz\, \vb, \\
\Ga^{\rho \, -1/2} &=& d\bz\, \vro, \\
\Ga^{\rho \, +1/2} &=& dz\, W^{\rho}_{3/2}
           + d\bz\, \Ga_\bz^{\rho \ +1/2}.
\ena
The zero curvature conditions yield, among other things
\be
\Ga_\bz^{\rho \ +1/2}
\ = \ \vz\, \wro + \vro\, W_1
		- \rho \, \dz\, \vro.
\ee

\indent

After transition to conformal parametrization, highest weight
constraints and projective pa\-ra\-me\-trization we are left with
a set of variables which are the pairs of $\cw$-gauge potentials
and currents, defined at lowest and highest weight,
respectively, for each value of $a$. More explicitly, for
$a=0$ we have, as usual,
\begin{center} $\vz \cem$ and $\cem \Lzz$, \end{center}
the \slt subsector with the projective connection. At $a=-1$ we have
\begin{center} $\vb \cem$ and $\cem W_1$, \end{center}
of conformal weights $(0,1)$ and $(1,0)$, respectively. Finally,
in the doublets at $a=-1/2$ with hypercharges $\rho=-1,+1$,
bosonic fields of half-integer conformal weights appear, namely
\begin{center} $\vro \cem$ and $\cem \wro$, \end{center}
of conformal weights $(-1/2,1)$ and $(3/2,0)$, respectively.

In terms of these quantities and after the recursive
procedure the zero curvature conditions then
take their final form as follows
\bea
\dbz \Lzz &=& \3 \vz - 2 W_1 \left(\vmi \, \wpl - \vpl \, \wmi \right)
\nn \\ [2mm]
    && - \vmi \, \dz \wpl -3 \wpl \, \dz \vmi
       - \vpl \, \dz \wmi -3 \wmi \, \dz \vpl, \\ [3mm]
\dbz W_1 &=& \dz \vb - \sm{3}{2} \left(\vmi \, \wpl	- \vpl \, \wmi
\right), \\ [3mm]
\dbz \wro &=& \vz \, \dz \wro + \sm{3}{2} \wro \, \dz \vz
                   - \rho \, \vz \, W_1 \, \wro + \rho \, \vb \, \wro
\nn \\ [2mm] && - \rho \left(\2 + W_1 W_1 \right) \vro	+ \vro \, \dz W_1
+ 2 W_1 \, dz \vro.
\ena
The ghost fields which survive at the end after the reduction procedure
are
\begin{itemize}
\item $c^z$, for the \slt sector, conformal weight $(-1,0)$,
\item $c$, for the hypercharge sector, conformal scalar,
\item $\cro$, for the two doublets of hypercharge $\rho=-1,+1$,
      conformal weights $(-1/2,0)$.
\end{itemize}

Having identified the basic variables in the gauge potential and the
ghost field sectors we turn now to their BRS differential algebra.
Following the general prescription we arrive at the BRS transformations
\bea
s \vz &=& \dbz c^z + c^z \, \dz \vz - \vz \, \dz c^z
    + \cmi \, \vpl -\vmi \, \cpl, \\ [3mm]
s \vb &=& \dbz c - 3 \left(\cmi \, \vpl - \vmi \, \cpl \right) \nn  \\
[2mm]  && + \sm{3}{2} \left( \cmi \, \dz \vpl - \vpl \, \dz \cmi \right)
+ \sm{3}{2} \left( \cpl \, \dz \vmi - \vmi \, \dz \cpl \right) \nn  \\
[2mm] && + \sm{3}{2} \ c^z \left(\vmi \, \wpl - \wmi \, \vpl \right)
	  + \sm{3}{2} \vz \left(\wmi \, \cpl - \cmi \, \wpl \right), \\ [3mm]
s \vro &=& \dbz \vro
+ c^z \, \dz \vro - \vz \, \dz \cro
+ \shalf \left(\cro \, \dz \vz - \vro  \, \dz c^z\right) \nn\\ [2mm]
&&+ \rho \left( c \, \vro - \vb \, \cro \right)
	+ \rho \, W_1 \left(\vz \, \cro - c^z \, \vro \right),
\ena
for the gauge potentials while those of the ghost fields are given as
\bea
s c^z &=& -c^z \, \dz c^z - \cmi \, \cpl, \\ [3mm]
s c &=& -\sm{3}{2} \left( \cpl \, \dz \cmi + \cmi \, \dz \cpl \right)
+ 3 W_1 \, \cmi \, \cpl \nn \\ [2mm]
&& + \sm{3}{2} c^z \left(\wmi \, \cpl - \cmi \, \wpl \right), \\ [3mm]
s \cro &=& - c^z \, \dz \cro - \shalf \, \cro \dz c^z
+ \rho \left(c^z W_1 - c \right) \cro.
\ena
Finally, for the $\cw$-currents, one arrives at
\bea
s \Lzz &=& \3 c^z - \cmi \, \dz \wpl - 3 \wpl \, \dz \cmi
   - \cpl \, \dz \wmi - 3 \wmi \, \dz \cpl \nn \\ [2mm]
&&+ 2 W_1 \left(\wmi \, \cpl - \cmi \, \wpl \right), \\ [3mm]
s W_1 &=& \dz c + \sm{3}{2} \left(\wmi \, \cpl - \cmi \, \wpl \right),
\\ [2mm] s \wro &=& - \rho \left(\2 + W_1 W_1 \right) \cro
+ c^z \, \dz \wro + \sm{3}{2} \wro \, \dz c^z \nn \\ [2mm]
&&	+ \cro \, \dz W_1 + 2 W_1 \, \dz \cro - \rho \left(c^z W_1 - c \right)
\wro.
\ena
The consistent anomaly, expressed in terms of the basic variables,
is a $(1,1)$ differential of ghost number one and hypercharge zero.
As before it is a particular combination of contributions from
the different ghosts:
\bea
\ca_{\bz z}^{(1)} &=& \left(c^z \dbz - \vz s \right) \Lzz
		- \sm{4}{3} \left(c \, \dbz - \vb s \right) W_1 \nn \\
&&  -2 \left(\cmi \dbz - \vmi s \right) \wpl
		-2 \left(\cpl \dbz - \vpl s \right) \wmi.
\ena
In some more detail this anomaly takes the following explicit form
\bea
\ca_{\bz z}^{(1)} &=&
c^z \3 \vz - \vz \3 c^z + \sm{4}{3} (\vb \, \dz c - c \, \dz \vb)
\nn \\[2mm]
&&+\ 2 \left(\cmi \, \2 \vpl + \vpl \, \2 \cmi
             - \cpl \, \2 \vmi - \vmi \, \2 \cpl \right) \nn \\[2mm]
&&+\ 4 W_1 \left( \vmi \dz \cpl - \cpl \dz \vmi
            + \vpl \dz \cmi - \cmi \dz \vpl \right) \nn \\[2mm]
&& +\ 3\wmi \left(\vpl \dz c^z - c^z \dz \vpl
            + \vz \dz \cpl - \cpl \dz \vz \right) \nn \\[2mm]
&&+\left(c^z \vpl - \vz \cpl\right) \dz \wmi \nn \\[2mm]
&& +\ 3\wpl \left(\vmi \dz c^z - c^z \dz \vmi
            + \vz \dz \cmi - \cmi \dz \vz \right) \nn \\[2mm]
&&+\left(c^z \vmi - \vz \cmi \right) \dz \wpl \nn \\[2mm]
&& +\ 4\wmi \left(\vb \, \cpl - c \, \vpl \right)
            +\ 4\wpl \left(c \, \vmi -\vb \, \cmi \right) \nn \\[2mm]
&&+\ 4 W_1 \wmi \left(c^z \vpl - \vz \cpl \right)
            -\ 4 W_1 \wpl \left(c^z \vmi - \vz \cmi \right),
\ena
where the terms are arranged such that conformal covariance becomes
as transparent as possible. Again, this expression can be obtained,
modulo trivial terms, from
\be \mbox{tr} \left. \left( \tld{A} \tld{d} \tld{A} \right) \right|_2^1,
\ee when expanded in the projective parametrization and subject to the
highest weight gauge, taking into account the second \slt decomposition
of this subsection in the evaluation of the trace.

\indent

Let us now come back to the discussion of the triplet
\be
\Sigma \ = \ \left(
\begin{array}{c}
\Sigma_{+1/2} \\ \Sigma_{-1/2} \\ \Sigma_0
\end{array}
\right).
\ee
The conformal parametrization,
\be \Psi \ = \ {}^{\hat{g}_0} \Sigma, \ee
is now obtained using the explicit matrix representation \rf{matr2}
of this subsection. As a consequence, we arrive at the redefinitions
\be
\psi_\zt\ =\ \sqrt{A_z^-} \, \Sigma_{+1/2}, \cem
\psi^\zt\ =\ \frac{1}{\sqrt{A_z^-}} \, \Sigma_{-1/2}, \cem
\psi\ =\ \Sigma_0,
\ee
introducing conformal weights $(+1/2,0)$, $(-1/2,0)$ and $(0,0)$,
respectively. The covariant derivatives on these components
in the conformal parametrization are
\bea
\cd \psi_\zt &=& (d-\sm{1}{2}\chi)\psi_\zt  + \lz \psi^\zt
	+ \Ga^{\ominus \ +1/2} \psi + \sm{1}{3} \Ga_Y \psi_\zt, \\
\cd \psi^\zt &=& (d+\sm{1}{2}\chi)\psi^\zt  - v^z \psi_\zt
	+ \Ga^{\ominus \ -1/2} \psi + \sm{1}{3} \Ga_Y \psi^\zt, \\
\cd \psi &=& d \psi + \Ga^{\oplus \ +1/2} \psi^\zt
	+ \Ga^{\oplus \ -1/2} \psi_\zt - \sm{2}{3} \Ga_Y \psi.
\ena
Imposing the constraint on $\Psi$ to be covariantly constant shows,
by virtue of the second equation, that $\psi_\zt$ becomes a
dependent variable and we are left, in the matter sector, with
the two basic fields $\psi^\zt$ and $\psi$. Taking into account
the projective and highest weight gauges, these basic fields are subject
to the conditions
\be
\dbz \psi^\zt \ = \  \vz \, \dz \psi^\zt - \shalf \psi^\zt \, \dz \vz
 + \sm{1}{3} \left(\vz W_1 - \vb \right) \psi^\zt - \vmi \psi,
\ee
\be
\left(\dz \dz + \shalf \Lzz \right) \psi^\zt
+ \sm{2}{3} \psi^\zt \, \dz W_1 + \sm{4}{3} W_1 \, \dz \psi^\zt
+ \wmi \, \psi \ =\ 0,
\ee
and
\be
\dbz \psi \ = \ - \vpl \, \dz \psi^\zt + \psi^\zt \, \dz \vpl
+\sm{2}{3} \vb \psi
  - \sm{4}{3} W_1 \, \vpl \, \psi^\zt - \vz \, \wpl \, \psi^\zt,
\ee
\be
\dz \psi - \sm{2}{3} W_1 \psi +  \wpl \, \psi^\zt \ = \ 0.
\ee
Finally, the BRS transformations of the basic variables in the matter
sector are obtained as
\bea
s \psi^\zt &=& c^z \, \dz \psi^\zt - \shalf \psi^\zt \, \dz c^z
	+ \sm{1}{3} (c^z W_1 - c) \psi^\zt - \cmi \, \psi, \\
s \psi &=& 	- \cpl \, \dz \psi^\zt + \psi^\zt \, \dz \cpl
+ \sm{2}{3} c \, \psi
 - \sm{4}{3} \cpl \,  W_1 \, \psi^\zt	- c^z \, \wpl \, \psi^\zt,
\ena
completing our presentation of the $\cw_3^{(2)}$-gauge structure.

\indent

\indent

\sect{SUMMARY AND CONCLUSIONS}

\indent

In this paper we have presented a procedure to identify what we have
called
$\cw$-gauge structures, starting from two dimensional flat Lie algebra
valued gauge potentials (and ghosts), with special emphasis on conformal
covariance properties.
The crucial point of our construction was a soldering process which,
for a given \slt decomposition of the corresponding Lie algebra and
in a certain highest weight parametrization, gave rise to
gauge structures in terms of a number of triplets
\be W_{a+2}, \cem v^{\ -a-1}_\bz, \cem c^{-a-1},  \ee
suggestively called $\cw$-currents (of conformal weight $(a+2,0)$),
$\cw$-gauge fields (of conformal weight $(-a-1,1)$), and
$\cw$-ghost fields (of conformal weight $(-a-1,0)$).
The values of $a$ occuring (\ie among $a \geq -1$, integer or
half-integer) and their multiplicities
depend on the properties of the particular \slt decomposition chosen.
In all cases there is, in addition to this set of triplets,
another one,
\be \Lzz, \cem \vz, \cem c^z, \ee
pertaining to the the \slt subsector, featuring the projective
connection $\Lzz$, crucial for conformal covariance.

The notion of gauge structure is justified by the action of the
nilpotent BRS antiderivation $s$ pertaining to the residual
$\cw$-gauge transformations, which is concisely defined on the
full set of currents, gauge and ghost fields.
It is instructive to discuss this BRS differential algebra
schematically.
For instance, the projective connection and the conformally
covariant currents transform as
\bea s \, \Lzz & \sim & \3 \ c^z \ \ \ + \ \ \ \mbox{diff. pol.}, \\
     s \, W_{a+2} & \sim &
        \Delta^{(2a+3)} \ c^{-a-1} \ \ \ + \ \ \ \mbox{diff. pol.},
\ena
exhibiting conformally covariant differential operators $\Delta^{(2a+3)}$
and {\em diff. pol.} standing for other conformally covariant
differential polynomials in terms of the basic fields, detailed
properties depending on the special case under consideration.
Observe that only $\prt$ derivatives (and no $\bp$ derivatives)
occur here. As to the BRS differential algebra, the $\bp$ derivatives
occur only in the transformations of the $\cw$-gauge fields
and there only linearly and in the particular combinations
\bea s \, \vz &-& \bp \, c^z \ \ \sim  \ \ \ \mbox{diff. pol.}, \\
     s \, v^{\ -a-1}_\bz &-& \bp \, c^{-a-1} \ \ \sim \ \ \ \mbox{diff.
pol.}.
\ena
This asymmetry between $\prt$ and $\bp$ derivatives manifests itself
also in the zero curvature conditions, appearing as certain
anholomorphicity conditions on the projective connection and
the currents, of the form
\bea \bp \, \Lzz & \sim & \3 \ \vz \ \ \ + \ \ \ \mbox{diff. pol.}, \\
     \bp \, W_{a+2} & \sim &
             \Delta^{(2a+3)} \ v^{\ -a-1}_\bz \ \ \ + \ \ \ \mbox{diff.
pol.},
\ena
very similar in structure to the above BRS transformations.
This qualitative presentation of the general case has been
illustrated in full detail for the examples of \slt and \sltr,
with special emphasis on covariantly chiral matter fields and
the anomaly structure.

As it stands, the procedure presented in this paper might be
understood as a purely algebraic algorithm which allows to derive
in a concise way the consistent BRS differential algebra
pertaining to $\cw$-geometry. A number of points like the explicit
construction of the highest weight parametrization (which was assumed
here as some constraint, consistent with the general structure,
and without any further justification) or the features of the covariant
constant matter fields in the general case deserve further study.
Likewise, possible dynamical realizations of the geometric concepts
presented here should be investigated. It should also be worthwile
to study possible relations with the quantum
Drinfeld-Sokolov reduction scheme as {\em e.g.} in \cite{dBT94}.

A more profound mathematical understanding might appeal,
at least in the \slt case, to concepts of
complex and projective structures in the context of flat complex vector
bundles over Riemann surfaces \cite{Gun67}, \cite{AB82}, \cite{Hit92}
for an interpretation of the soldering procedure in terms of
a special representative for the corresponding connections. The
projective parametrization in such a picture might be related to the
change of complex structure on the Riemann surface through smooth
diffeomorphisms, with $\vz$ playing eventually the role of a Beltrami
differential.

\end{document}